\documentclass[ALICE,manyauthors]{cernphprep}

\newcommand{\state}{$\Upsilon$(1S) }
\newcommand{\stateP}{$\Upsilon$(2S) }
\newcommand{\statePP}{$\Upsilon$(3S) }
\newcommand{\raa}{$R_{\rm AA}$ }

\newcommand{\sNN}{$\sqrt{s_{\rm NN}}=2.76\,{\rm TeV}$ }

\usepackage{rotating}
\usepackage{hyperref}
\usepackage{xcolor}
\usepackage{doi}
\usepackage[sort&compress,numbers]{natbib}
\usepackage{lineno}

\begin{document}%

\begin{titlepage}
\PHyear{2014}
\PHnumber{103}      
\PHdate{17 May}  
%

\title{Suppression of \state at forward rapidity \\in Pb\---Pb collisions at $\mathbf{ \sqrt{{\it s}_{\mathbf NN}} = 2.76}$ TeV}

\ShortTitle{\state at forward rapidity in Pb\--Pb collisions \hbox{at \sNN}}

\Collaboration{ALICE Collaboration\thanks{See Appendix~\ref{app:collab} for the list of collaboration members}}
\ShortAuthor{ALICE Collaboration} 

\begin{abstract}
We report on the measurement of the inclusive \state production in Pb\---Pb collisions at $\sqrt{s_{\rm NN}}=2.76$~TeV carried out at forward rapidity ($2.5<y<4$) and down to zero transverse momentum using its $\mu^{+}\mu^{-}$ decay channel with the ALICE detector at the Large Hadron Collider. A strong suppression of the inclusive \state yield is observed with respect to pp collisions scaled by the number of independent nucleon-nucleon collisions.
The nuclear modification factor, for events in the 0\---90\% centrality range, amounts to $0.30\pm0.05{\rm (stat)}\pm0.04{\rm (syst)}$. The observed \state suppression tends to increase with the centrality of the collision and seems more pronounced than in corresponding mid-rapidity measurements. Our results are compared with model calculations, which are found to underestimate the measured suppression and fail to reproduce its rapidity dependence.
\end{abstract}
\end{titlepage}
\setcounter{page}{2}

\section{Introduction}
\label{Introduction}
\raggedbottom
At high temperature and high density, Quantum Chromodynamics predicts the existence of a deconfined state of strongly-interacting matter (Quark-Gluon Plasma, QGP) with properties governed by the quark and gluon degrees of freedom~\cite{Shuryak:1980tp}. This state can be studied in ultra-relativistic heavy-ion collisions and is expected to be produced when the temperature of the system exceeds the critical temperature $T_{\rm c}\simeq150-195~{\rm MeV}$~\cite{Cheng:2009zi,Borsanyi:2010bp}. Among the particles which can be measured to investigate the QGP properties, heavy quarks are of special interest since they are produced in the initial parton-parton interactions and they interact with the medium throughout its evolution. In particular, the study of the heavy quark-antiquark bound state (quarkonium) is expected to provide essential information on QGP properties. The colour-screening model~\cite{Matsui:1986dk} predicts that charmonia and bottomonia ($c\overline{c}$ and $b\overline{b}$ bound states, respectively) dissociate in the medium, resulting in a suppression of the observed yields. More specifically, the quarkonium binding properties are expected to be modified in the deconfined medium and, out of the various charmonium and bottomonium states, the less tightly bound might melt close to $T_{\rm c}$ and the most tightly bound well above $T_{\rm c}$~\cite{Digal:2001ue}. A sequential suppression pattern with increasing temperature is then expected to be realized. Based on results from quenched lattice QCD~\cite{Wong:2004zr,Satz:2005hx}, the most tightly bound bottomonium state, $\Upsilon{\rm (1S)}$, is predicted to melt at a temperature larger than $4~T_{\rm c}$, while the $\Upsilon$(2S) and the $\Upsilon$(3S) should melt at $1.6$ and $1.2~T_{\rm c}$, respectively. The melting temperature for the J/$\psi$ charmonium state is expected to be close to that of the \stateP and the \statePP bottomonium states. 
In the case of recent spectral-function approaches with complex potential~\cite{Mocsy:2007jz,Petreczky:2010tk}, the obtained dissociation temperatures are lower.

In the charmonium sector, a significant suppression of the J/$\psi$ yield has been observed at SPS~\cite{Abreu:1998wx,Alessandro:2004ap,Arnaldi:2007zz} ($\sqrt{s_{\rm NN}}=17.3\,{\rm GeV}$), RHIC~\cite{Adare:2006ns,Abelev:2009qaa} ($\sqrt{s_{\rm NN}}=39$, $62.4$, $200\,{\rm GeV}$) and LHC~\cite{Abelev:2012rv,Aad:2010aa,Chatrchyan:2012np} ($\sqrt{s_{\rm NN}}=2.76\,{\rm TeV}$) energies.
A qualitative description of the results can be
obtained assuming that in addition to the dissociation by colour screening, a
regeneration process takes place for high-energy collisions. 
The regeneration mechanism is particularly important at LHC energies, where the multiplicity of charm quarks is large~\cite{Grandchamp:2003uw,Bratkovskaya:2004cq,Thews:2000rj,BraunMunzinger:2000px,Andronic:2003zv}. 
The $\psi$(2S) charmonium state has lower binding energy than the J/$\psi$ one and cannot be produced by the decays of higher mass states. At SPS energies~\cite{Alessandro:2006ju}, the suppression of $\psi$(2S) yield is about $2.5$ times larger than for the J/$\psi$ state.
With the high collision energies and luminosities recently available at RHIC and LHC, it is also possible to study bottomonium production in heavy-ion collisions~\cite{Adamczyk:2013poh,PHENIXUps,Chatrchyan:2011pe,Chatrchyan:2012lxa,PN}.  
Compared with the J/$\psi$ case, the probability for the $\Upsilon$ states to be regenerated in the medium is much smaller due to the lower production cross section of $b\overline{b}$ pairs~\cite{RefEmRap}.  
However, the feed-down from higher mass bottomonia (between $40\%$ and $50\%$ for $\Upsilon$(1S)~\cite{QuarkoppALICE}) complicates the data interpretation. Furthermore, the suppression due to the QGP must be disentangled from that due to Cold Nuclear Matter (CNM) effects (such as nuclear modification of the parton distribution functions or break-up of the quarkonium state in CNM) which, as of now, are not accurately known neither at RHIC energies~\cite{Adamczyk:2013poh} 
nor in the forward rapidity regions probed at LHC. At RHIC, the inclusive $\Upsilon$(1S+2S+3S) production has been measured in Au\--Au collisions at mid-rapidity by the STAR~\cite{Adamczyk:2013poh} and PHENIX~\cite{PHENIXUps} Collaborations. The observed suppression is consistent with the melting of the \stateP and \statePP states. At LHC, the CMS Collaboration has measured the mid-rapidity production of bottomonium states in Pb\---Pb collisions. The \state yield is suppressed by approximately a factor of two with respect to the expectation from pp collisions obtained by scaling of the hard process yield with the number of binary nucleon-nucleon collisions. Moreover, the \stateP and the \statePP are almost completely suppressed~\cite{Chatrchyan:2011pe,Chatrchyan:2012lxa}.

In this Letter, we report on the inclusive \state production at forward rapidity ($2.5<y<4$) and down to zero transverse momentum ($p_{\rm T}>0$) in Pb\---Pb collisions at $\sqrt{s_{\rm NN}}=2.76\,{\rm TeV}$. The measurement was carried out in the $\mu^{+}\mu^{-}$ decay channel with the ALICE detector. The yield of \state in Pb\---Pb collisions relative to pp, normalized to the number of nucleon-nucleon collisions at the same energy (nuclear modification factor, $R_{\rm AA}$) is reported in two centrality intervals and two rapidity intervals. The results are compared with CMS \state mid-rapidity data~\cite{Chatrchyan:2012lxa} and with model calculations~\cite{Emerick:2011xu,Strickland:2012cq}.

\section{Experimental apparatus and data sample}

The ALICE detector is described in detail in reference~\cite{Aamodt:2008zz}. In this Section, we briefly summarize the main features of the detectors used for this analysis.  
The muon spectrometer, covering a pseudo-rapidity range $-4<\eta_{\rm lab}<-2.5$ in the laboratory frame\footnote{In the ALICE reference frame, the positive z-direction is along the counter clockwise beam direction. Thus, the muon spectrometer covers a negative pseudorapidity ($\eta_{\rm lab}$) range and a negative $y$ range. In this Letter the results are presented with a positive $y$ notation keeping the $\eta_{\rm lab}$ values signed.}, consists primarily of a tracking apparatus composed of five stations of two planes of Cathode Pad Chambers (CPC) each, a dipole magnet delivering a 3 T$\cdot$m integrated magnetic field used to bend the charged particles in the tracking system area and a triggering system including four planes of Resistive Plate Chambers (RPC). The detector incorporates a 10 interaction length front absorber used to filter the muons upstream of the tracking apparatus and a 7.2 interaction length iron wall located between the tracking and the triggering systems. The iron wall plays an important role in the muon identification, since it stops the light hadrons escaping from the front absorber and the low momentum background muons produced mainly in $\pi$ and K decays. 

The V0 detector~\cite{VZEROJINST} consists of two scintillator arrays covering the full azimuth and the pseudo-rapidity ranges $2.8<\eta_{\rm lab}<5.1$ (V0-A) and $-3.7<\eta_{\rm lab}<-1.7$ (V0-C). Both scintillator arrays have an intrinsic time resolution better than 0.5 ns~\cite{VZEROJINST,ALICEPerfPaper} and their timing information was used for offline rejection of events produced by the interactions of the beam with residual gas (or beam-gas interactions). 

The Zero Degree Calorimeters (ZDC), which are located at $114$ meters on each side of the ALICE interaction point, were used to reduce the beam-halo background by means of an offline timing cut~\cite{ALICEPerfPaper}.  
Another cut on the energy deposited in the ZDC suppresses the background contribution from electromagnetic Pb\---Pb interactions.

Finally, the Silicon Pixel Detector (SPD) is used to reconstruct the primary vertex. This detector consists of two cylindrical layers covering the full azimuth and the pseudo-rapidity ranges $|\eta|<2.0$ and $|\eta|<1.4$ for the inner and outer layer, respectively.

The Minimum-Bias (MB) trigger is defined as the coincidence of a signal in the two V0 arrays. The efficiency of such a trigger for selecting inelastic Pb\---Pb interactions is larger than $95\%$~\cite{Abelev:2013qoq}. 
In order to enrich the data sample with dimuons, the trigger used in this analysis requires the detection of an opposite-sign muon pair in the triggering system in coincidence with the MB condition. The muon trigger system selects tracks having a transverse momentum, $p_{\rm T}^{\mu}$, larger than $1\,{\rm GeV/}c$. This threshold is not sharp and the quoted value corresponds to a 50\% trigger probability on a muon candidate. Events were classified according to their degree of centrality, which is calculated through the study of the V0 signal amplitude distribution~\cite{Aamodt:2010cz}. This analysis was carried out for the events corresponding to the most central $90\%$ of the inelastic Pb\---Pb cross section. In this centrality range, the efficiency of the MB trigger for selecting inelastic Pb\---Pb interactions is $100\%$ and the contamination from electromagnetic processes is negligible. The analysed data sample corresponds to an integrated luminosity $L_{\rm int}=68.8\pm0.9{\rm (stat)}^{+6.0}_{-5.1}{\rm (syst)}\,\mu{\rm b^{-1}}$~\cite{Abelev:2013ila}.

\section{Data analysis}

\begin{figure}[thb!f]
\begin{center}
\includegraphics[width=7.5cm]{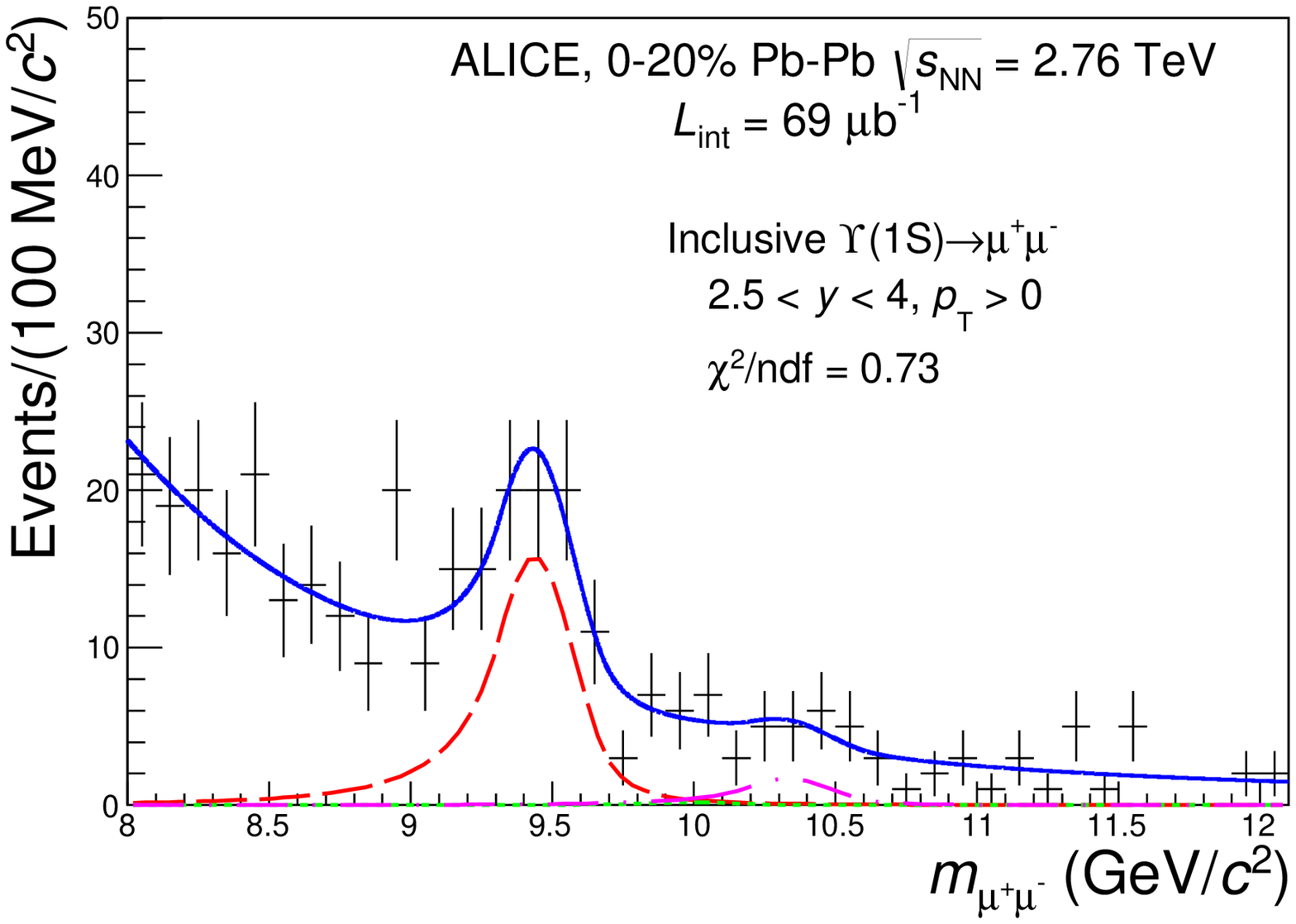}
\includegraphics[width=7.5cm]{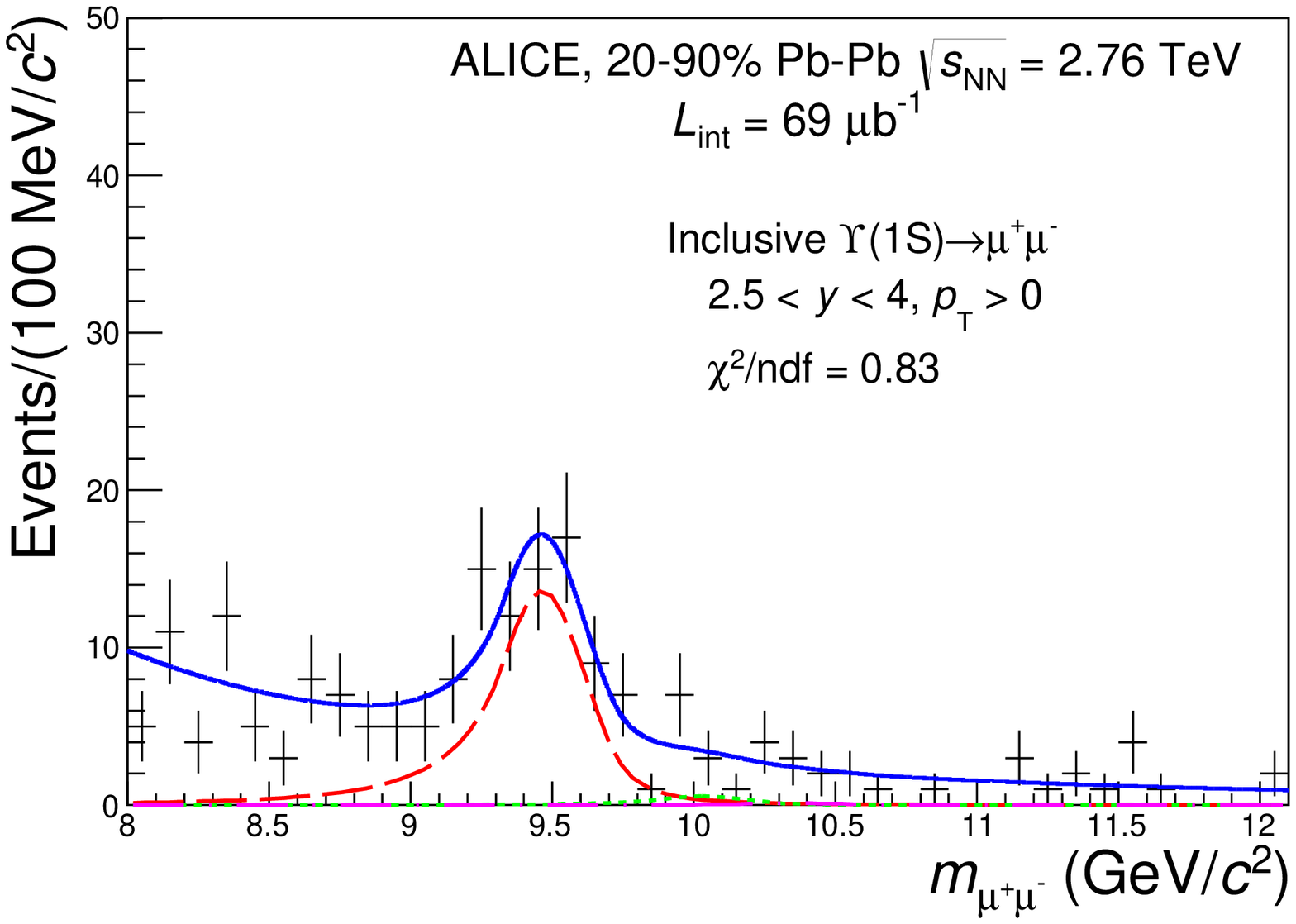}
\includegraphics[width=7.5cm]{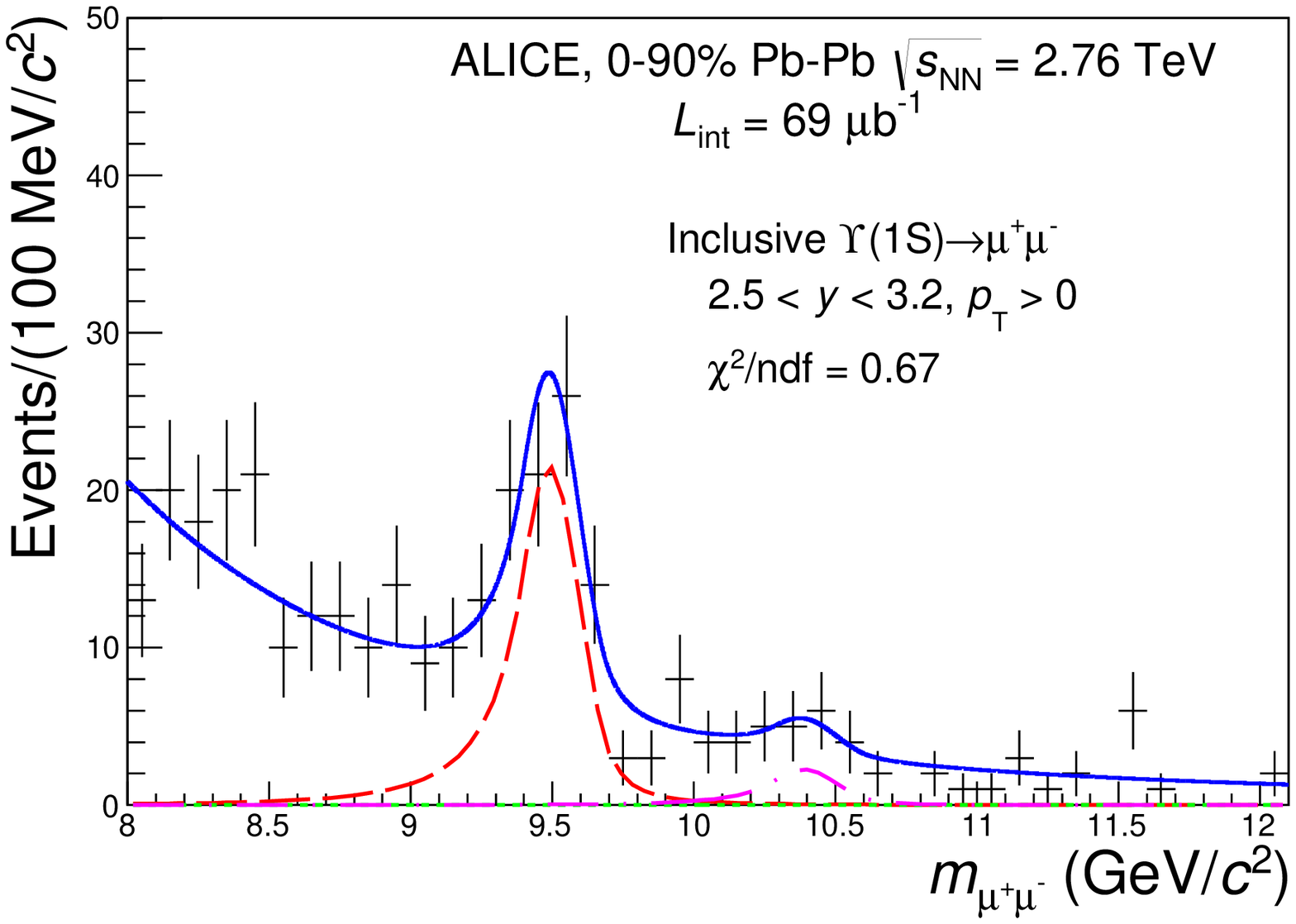}
\includegraphics[width=7.5cm]{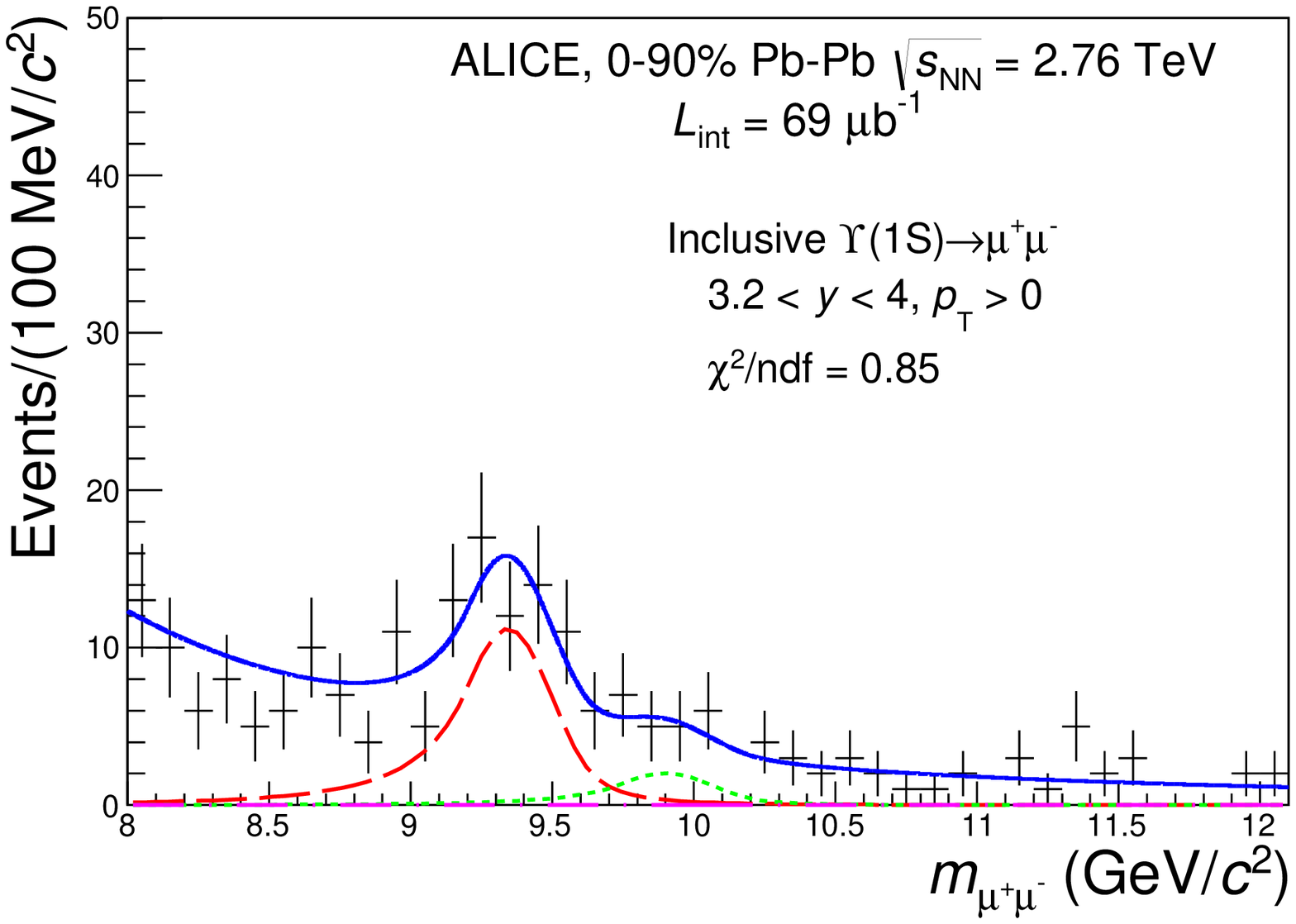}
\includegraphics[width=7.5cm]{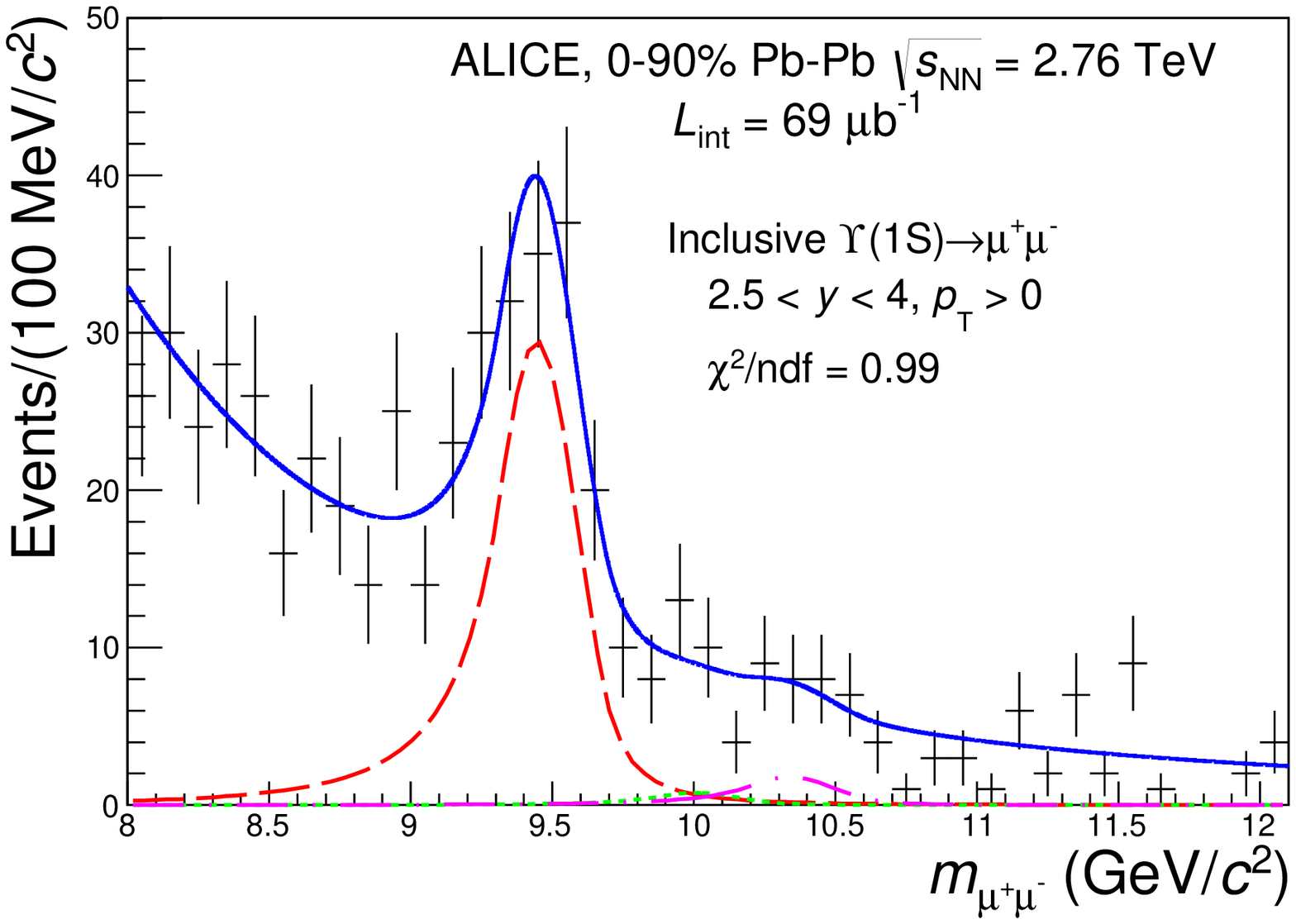}
\caption{Invariant mass distribution of opposite-sign dimuons with $p_{\rm T}>0$ for the different centrality and rapidity intervals considered in the analysis (see text for details). The solid blue line represents the total fit function (sum of two exponential and three extended Crystal Ball functions) and the dashed red line is the \state signal component only. The green dotted line and the magenta dashed-dotted line represent the \stateP and the \statePP peaks, respectively.}
\label{FitFig}
\end{center}
\end{figure}

Several steps are necessary to estimate the \state nuclear modification factor. They are described in the following section. Additional details on the analysis can be found in~\cite{PN}.

Muon track candidates were reconstructed starting from the hits in the tracking chambers~\cite{Aamodt:2011gj}. Each reconstructed track was then required to match a track segment in the trigger chambers (trigger tracklet) and to have a transverse momentum $p_{\rm T}^{\mu}>2\,{\rm GeV/}c$. The latter requirement helps in reducing the contribution of soft muons from $\pi$/K decays without affecting muons from \state decays. A further selection was applied by requiring the muon tracks to exit the front absorber at a radial distance from the beam axis, $R_{\rm abs}$, in the range $17.6<R_{\rm abs}<89.5$ cm. This selection rejects tracks crossing the region of the absorber with the material of highest density, where multiple-scattering and energy-loss effects are large and affect the mass resolution. Finally, each track was required to point to the interaction vertex in order to reject the contributions from fake tracks and beam-gas interactions.
Tracks were then combined to form opposite-sign muon pairs and a $2.5<y<4$ cut on the pair rapidity was introduced to remove dimuons at the edge of the acceptance.

The raw number of \state was obtained by means of a fit to the dimuon invariant mass distributions with the combination of several functions (see Fig.~\ref{FitFig}). The background was parametrized as the sum of two exponential functions with all parameters let free. Such functions reproduce well the data on the large invariant mass range of our fits, 5\--18 GeV/$c^2$. Monte Carlo simulations show that each $\Upsilon$ resonance shape is well described by an extended Crystal Ball (CB) function~\cite{CB2} made of a Gaussian core and a power-law tail on both sides. The low invariant mass tail is due to non-Gaussian multiple scattering in the front absorber, while the high invariant mass one is due to alignment and calibration biases. In the fit, the position and the width of the \state peak were left free, as they can be constrained by the data themselves. The position of the \stateP and \statePP peaks were fixed to that of the \state according to the PDG~\cite{PDGTabRef} mass difference, while their widths were forced to scale 
proportionally to that of the \state according to the ratio of the resonance masses. This scaling was verified to be fulfilled in MC simulations. The CB tails are poorly constrained by the data and were fixed using MC simulations.   
Fits were performed on the $y$-integrated, 0\---90\% centrality distribution, as well as for two centrality intervals, 0\---20\% (central collisions) and 20\---90\% (semi-peripheral collisions), or two rapidity ranges, $2.5<y<3.2$ and $3.2<y<4$. The tail parameters depend on rapidity but remain constant with respect to centrality. 
For each of the mentioned intervals, the significance (S/$\sqrt{{\mathrm S}+{\mathrm B}}$), evaluated on a range centered on the $\Upsilon$(1S) peak position and ranging between $\pm3$ times its width, is larger than five and the signal-to-background ratio larger than one. In the case of the \stateP and $\Upsilon$(3S), the significance and the signal-to-background ratio are too low to separate the signal from the underlying background.
The \state mass, as extracted from the fit, is consistent with the resonance mass value from the PDG~\cite{PDGTabRef}. Depending on the considered rapidity range, its width ranges from $(107\pm25)\,{\rm MeV/}c^{2}$ to $(159\pm40)\,{\rm MeV/}c^{2}$ and is consistent with the results from MC simulations. 

In order to estimate the systematic uncertainties on the signal extraction, the fits were performed over several invariant mass ranges and a sum of two power-law functions was used as an alternative parametrization of the background. Concerning the resonance peaks, alternative choices were made for the values of the fit parameters that were kept fixed in the default procedure outlined above. First, the width and the position of the \stateP and \statePP were 
varied by an amount corresponding to the size of the uncertainties on the corresponding fit parameters for the $\Upsilon$(1S). Then, the CB tail parameters were varied according to the uncertainties in their determination from fits of the MC signal distributions. 
For each source of systematic uncertainty (background parametrization, fixed widths and positions as well as tail parameters), the Root Mean Square (RMS) of the distribution of signal counts obtained with the different fits was estimated and the corresponding relative uncertainties were summed in quadrature.

With these prescriptions the number of \state counts is $134\pm 20{\rm (stat)}\pm 7{\rm (syst)}$ in the rapidity range $2.5<y<4$ and 0\---90\% centrality. Depending on centrality and rapidity, the systematic uncertainties range between $5\%$ and $10\%$. They are almost constant with centrality and reach a maximum in the $3.2<y<4$ rapidity interval.

The measured number of \state was corrected for the detector acceptance and efficiency ($A\times\varepsilon$) estimated by means of an Embedding Monte Carlo (EMC) method. The MC hits of muons from \state decays were embedded into MB events at the raw-data level. The standard reconstruction algorithm~\cite{Aamodt:2011gj} was then applied to these events. This method reproduces the detector response to the signal in a highly realistic background environment and accounts for possible variations of the reconstruction efficiency with the collision centrality.  
The $p_{\rm T}$ and $y$ distributions of the generated \state were obtained from existing pp measurements~\cite{Acosta:2001gv,LHCb:2012aa,Khachatryan:2010zg} using the extrapolation procedure described in~\cite{ParamUpSystRef}. EKS98 nuclear shadowing calculations~\cite{Eskola:1998df} were used to include an estimate of CNM effects. Since available data favour a small or null polarization for $\Upsilon$(1S)~\cite{Abazov:2008aa,CDF:2011ag,Chatrchyan:2012woa}, an unpolarised production was assumed (in both pp and Pb\---Pb collisions). The variations of the performance of the tracking and triggering systems throughout the data-taking period as well as the residual misalignment of the tracking chambers were taken into account in the EMC.

Four contributions enter the systematic uncertainty on $A\times\varepsilon$: (i) the input \state $p_{\rm T}$ and $y$ distributions for EMC, (ii) the tracking efficiency, (iii) the trigger efficiency and (iv) the matching of trigger tracklets with tracks in the tracking system. Type (i) uncertainties correspond to the maximum difference between $A\times\varepsilon$ evaluated by using the default input parametrizations and those obtained by using parametrizations corresponding to pp and Pb\---Pb collisions at different energies and centralities. The tracking and trigger efficiencies determined from data~\cite{Aamodt:2011gj} and from MC simulations were compared to evaluate type (ii) and (iii) contributions. For the type (iv) systematic uncertainties, the estimate was performed by varying by a similar amount, in both MC and real data, the value of the $\chi^{2}$ cut of the matching probability between reconstructed tracks in the tracking system and trigger tracklets. The comparison of the results of the two approaches provides the uncertainty. 

For \state produced in $2.5<y<4$ with $p_{\rm T}>0$, the value of $A\times\varepsilon$ is $0.226\pm0.025 {\rm (syst)}$ in semi-peripheral collisions and decreases to $0.216\pm0.024 {\rm (syst)}$ in central collisions. For the centrality-integrated sample the value of $A\times\varepsilon$ is $0.219\pm0.024 {\rm (syst)}$. Depending on centrality and rapidity, the systematic uncertainties range between $11\%$ and $12\%$.

The raw number of $\Upsilon$(1S), $N[\Upsilon{\rm (1S)}]$, was corrected for the acceptance and efficiency, $(A\times\varepsilon)$, and for the branching ratio of the dimuon decay channel, ${\rm BR}_{\Upsilon{\rm(1S)}\rightarrow\mu^{+}\mu^{-}} = 0.0248\pm0.0005$~\cite{PDGTabRef}. The yield, $Y_{\Upsilon{\rm (1S)}}$, was then obtained by normalizing the result to the equivalent number of MB events, $N_{\rm MB}$, via

\begin{center}
\begin{equation}
Y_{\Upsilon{\rm (1S)}}~=~\frac{N[\Upsilon{\rm (1S)}]}{(A\times\varepsilon)\times{\rm BR}_{\Upsilon{\rm (1S)}\rightarrow\mu^{+}\mu^{-}}\times N_{\rm MB}}.
\label{YieldFormula}
\end{equation}
\end{center}
  
\noindent Since the analysis is based on a dimuon trigger sample, the equivalent number of MB events was obtained by multiplying the number of triggered events by an enhancement factor, $F$, which corresponds to the inverse of the probability of having the dimuon trigger condition verified in an MB event. The $F$ factor averaged over the data-taking period is $F=27.5\pm1.0{\rm (syst)}$, where the systematic uncertainty reflects the spread of its values observed in the different periods of data taking. Within the rapidity interval $2.5<y<4$, the $\Upsilon$(1S) yield is $Y_{\Upsilon{\rm (1S)}}=(5.2\pm0.8 {\rm (stat)}\pm0.7 {\rm (syst)})\times10^{-5}$. The values of the yields in the other centrality and rapidity ranges considered in the analysis are given in Table~\ref{Yieldvalues}.

\begin{table}[h]
\begin{center}
\begin{tabular}{ccc}
Centrality& Rapidity & $({\rm Yield}\pm{\rm stat}\pm{\rm uncorr}\pm{\rm corr})\times10^{5}$\\
\hline
\hline
0\---20\% & $2.5<y<4$ & $11.3\pm2.5\pm0.7\pm1.3$ \\
20\---90\%& $2.5<y<4$  & $3.2\pm0.6\pm0.2\pm0.4$ \\
0\---90\% & $2.5<y<3.2$ & $3.2\pm0.6\pm0.4\pm0.1$ \\
0\---90\%& $3.2<y<4$  & $1.9\pm0.4\pm0.3\pm0.1$ \\
\end{tabular}
\end{center}
\caption{\label{Yieldvalues} Yields for the different centrality and rapidity intervals considered in the analysis. Statistical uncertainties are referred to as stat, uncorrelated systematic uncertainties as uncorr and correlated systematic uncertainties as corr. When results are integrated on rapidity (centrality), the degree of correlation is mentioned with respect to centrality (rapidity).}
\end{table}

The medium effects on the yields can be quantified by means of the nuclear modification factor

\begin{center}
\begin{equation}
R_{\rm AA}=\frac{Y_{\Upsilon{\rm (1S)}}}{\langle T_{\rm AA}\rangle\times\sigma^{\rm pp}_{\Upsilon{\rm (1S)}}},
\label{NuclModFactFormula}
\end{equation}
\end{center}

\noindent where $\langle T_{\rm AA}\rangle$ is the average nuclear overlap function, which can be interpreted as the average number of nucleon-nucleon binary collisions normalized to the inelastic nucleon-nucleon cross section, and $\sigma^{\rm pp}_{\Upsilon{\rm (1S)}}$ is the \state production cross section in pp collisions at $\sqrt{s}=2.76\,{\rm TeV}$.

The number of participant nucleons, $\langle N_{\rm part}\rangle$, and the $\langle T_{\rm AA}\rangle$ corresponding to each centrality class used in this analysis were obtained from a Glauber model calculation~\cite{Abelev:2013qoq}. Table~\ref{GlaubTab} shows the correspondence between the centrality class, $\langle N_{\rm part}\rangle$ and $\langle T_{\rm AA}\rangle$. The average number of participant nucleons weighted by the number of binary nucleon-nucleon collisions, $\langle N_{\rm part}^{\rm w}\rangle$, is also shown. The weighted average was calculated for each centrality class according to the values reported in~\cite{Abelev:2013qoq} for narrow intervals. The $\langle N_{\rm part}^{\rm w}\rangle$ quantity represents a more precise evaluation of the average centrality for a given interval, since the \state production is a hard process
and its initial yield scales with the number of binary nucleon-nucleon collisions, in the absence of initial-state effects.

\begin{table}[h]
\begin{center}
\begin{tabular}{cccc}
Centrality&$\langle N_{\rm part}\rangle$ & $\langle N_{\rm part}^{\rm w}\rangle$  & $\langle T_{\rm AA}\rangle$ (mb$^{-1}$)\\
\hline
\hline
0\---90\% & $124\pm2$ & $262\pm4$ &$6.3\pm0.2$\\
0\---20\% & $308\pm5$ & $323\pm5$ & $18.9\pm0.6$ \\
20\---90\%& $72\pm3$ & $140\pm6$ & $2.7\pm0.1$ \\
\end{tabular}
\end{center}
\caption{\label{GlaubTab} Correspondence between the centrality class, the average number of participant nucleons $\langle N_{\rm part}\rangle$, the average number of participant nucleons weighted by the number of binary nucleon-nucleon collisions $\langle N_{\rm part}^{\rm w}\rangle$, and the average nuclear overlap function $\langle T_{\rm AA}\rangle$. The values are obtained as described in~\cite{Abelev:2013qoq}.}
\end{table}

\begin{table}[h]
\begin{center}
\begin{tabular}{lccc}
\textrm{Source}&
\textrm{Centrality}&
\multicolumn{1}{c}{\textrm{Rapidity}} & \textrm{Integrated}\\
\hline
\hline
\textrm{signal extraction} & 5\---6\%(II) & 5\---10\%(II) & 5\%\\
\textrm{input EMC distributions} &  4\%(I) & 5\---7\%(II) & 4\%\\
\textrm{tracking efficiency}& 10\%(I) & 9\---11\%(II) & 10\%\\
\textrm{trigger efficiency}& 2\%(I) & 2\%(II) & 2\%\\
\textrm{matching efficiency}& 1\%(I)& 1\%(II) &  1\%\\
$\langle T_{\rm AA}\rangle$& 3\---4\%(II) & 3\%(I) & 3\%\\
$N_{\rm MB}$& 4\%(I) & 4\%(I) &4\%\\
${\rm BR}_{\Upsilon{\rm(1S)}\rightarrow\mu^{+}\mu^{-}}\times\sigma^{\rm pp}_{\Upsilon{\rm (1S)}}$ & 4\%(I) & 4\---7\%(II) 4\%(I)& 4\%\\
\end{tabular}
\end{center}
\caption{\label{tab:table1} Summary of the relative systematic uncertainties on each quantity entering the \raa calculation for centrality and rapidity ranges. The type I (II) stands for correlated (uncorrelated) uncertainties. When two values are given for type II uncertainties, the first value is given for the 0\---20\% ($2.5<y<3.2$) centrality (rapidity) interval, the second one for the 20\---90\% ($3.2<y<4$) interval. The values of systematic uncertainties for the $R_{AA}$ integrated over 0\---90\% in centrality and $2.5<y<4$ in rapidity are quoted in the last column.}
\end{table}

Due to the limited number of events collected in pp collisions at $\sqrt{s}=2.76\,{\rm TeV}$, we cannot measure $\sigma^{\rm pp}_{\Upsilon{\rm (1S)}}$. Instead, the LHCb data~\cite{Aaij:2014nwa} are used for the $R_{\rm AA}$ estimate\footnote{When ALICE preliminary results were released, the LHCb data were not yet available and $\sigma^{\rm pp}_{\Upsilon{\rm (1S)}}$ was estimated using a data-driven method as explained in~\cite{PN}. Depending on the rapidity interval, the pp reference obtained with this approach and the LHCb data~\cite{Aaij:2014nwa} differ by 30\---35\%. Taking into account uncertainties, it implies a change on the modification factor by $1.3$ to $2.2\sigma$, depending on rapidity.}.
LHCb quotes $\sigma^{\rm pp}_{\Upsilon{\rm (1S)}}\times{\rm BR}_{\Upsilon{\rm(1S)}\rightarrow\mu^{+}\mu^{-}}=0.670\pm0.025{\rm (stat)}\pm0.026{\rm (syst)}$ nb in the $2.5<y<4$ rapidity range. For the rapidity intervals studied in this analysis ($2.5<y<3.2$ and $3.2<y<4$) there is no exact matching with the rapidity ranges provided by LHCb. Therefore, a rapidity interpolation was performed to provide the values corresponding to our intervals. The LHCb data, with the statistical and uncorrelated systematic uncertainties summed in quadrature, were fitted with Gaussian or 
even-degree polynomial functions. The functions were then integrated 
over the required rapidity region and, for each range, the \state pp 
cross-section result is the average of the values obtained with the 
various fitting functions. The associated uncorrelated systematic uncertainty is obtained summing in quadrature the largest fit uncertainty and the half spread of the 
results obtained with the different fitting functions. The correlated 
systematic uncertainty associated to the LHCb values is taken as a 
further correlated contribution to the uncertainty of our interpolation 
result. More details on the pp reference are given in~\cite{PN}.

The relative systematic uncertainties on each quantity entering the \raa calculation are listed in Table~\ref{tab:table1}. 

\begin{table}[h]
\begin{center}
\begin{tabular}{ccc}
\hline
 Centrality &  Rapidity & {\rm $R_{\rm AA}\pm{\rm stat}\pm{\rm uncorr}\pm{\rm corr}$}\\
\hline
\hline
 0\---20\%& $2.5<y<4$& $0.22\pm 0.05\pm 0.02\pm0.03$ \\
 20\---90\%& $2.5<y<4$ & $0.44\pm 0.09\pm0.03\pm0.05$ \\
0\---90\%  & $2.5<y<3.2$& $0.30\pm 0.05\pm0.04\pm0.02$ \\
0\---90\%  &$3.2<y<4$& $0.29\pm 0.07\pm0.05\pm0.02$
\end{tabular}
\end{center}
\caption{\label{ValueRAA} Values of the \raa measured in the centrality and rapidity ranges considered in this analysis. Statistical uncertainties are referred to as stat, uncorrelated systematic uncertainties are referred to as uncorr and correlated systematic uncertainties are referred to as corr.}
\end{table}

\begin{figure}[h*]
\begin{center}
\includegraphics[width=11cm]{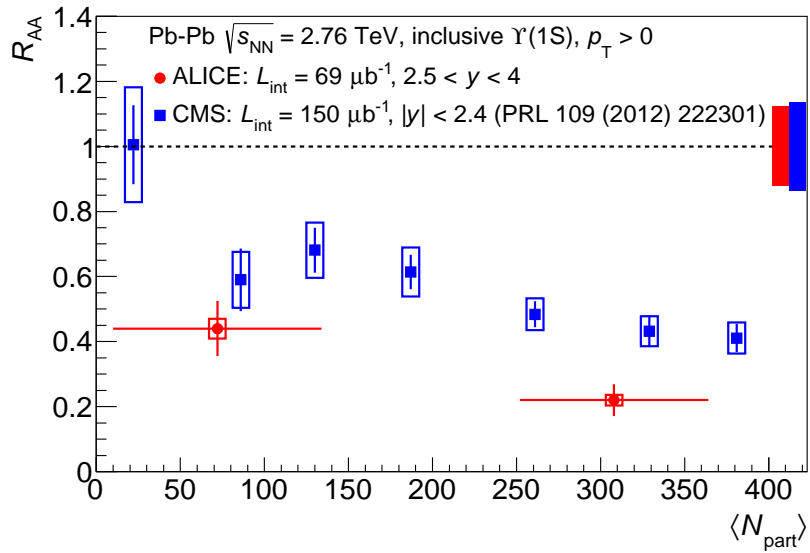}
\caption{Inclusive \state \raa as a function of the average number of participant nucleons. ALICE data refer to the rapidity range $2.5<y<4$ and are shown together with CMS~\cite{Chatrchyan:2012lxa} data which are reported in $|y|<2.4$. The vertical bars represent the statistical uncertainties and the boxes the point-to-point uncorrelated systematic uncertainties. The relative correlated uncertainties (12\% for ALICE and 14\% for CMS) are shown as a box at unity. The point-to-point horizontal error bars correspond to the RMS of the $N_{\rm part}$ distribution.}
\label{ALICEJpsiFig}
\end{center}
\end{figure}

\begin{figure}[h*]
\begin{center}
\includegraphics[width=11cm]{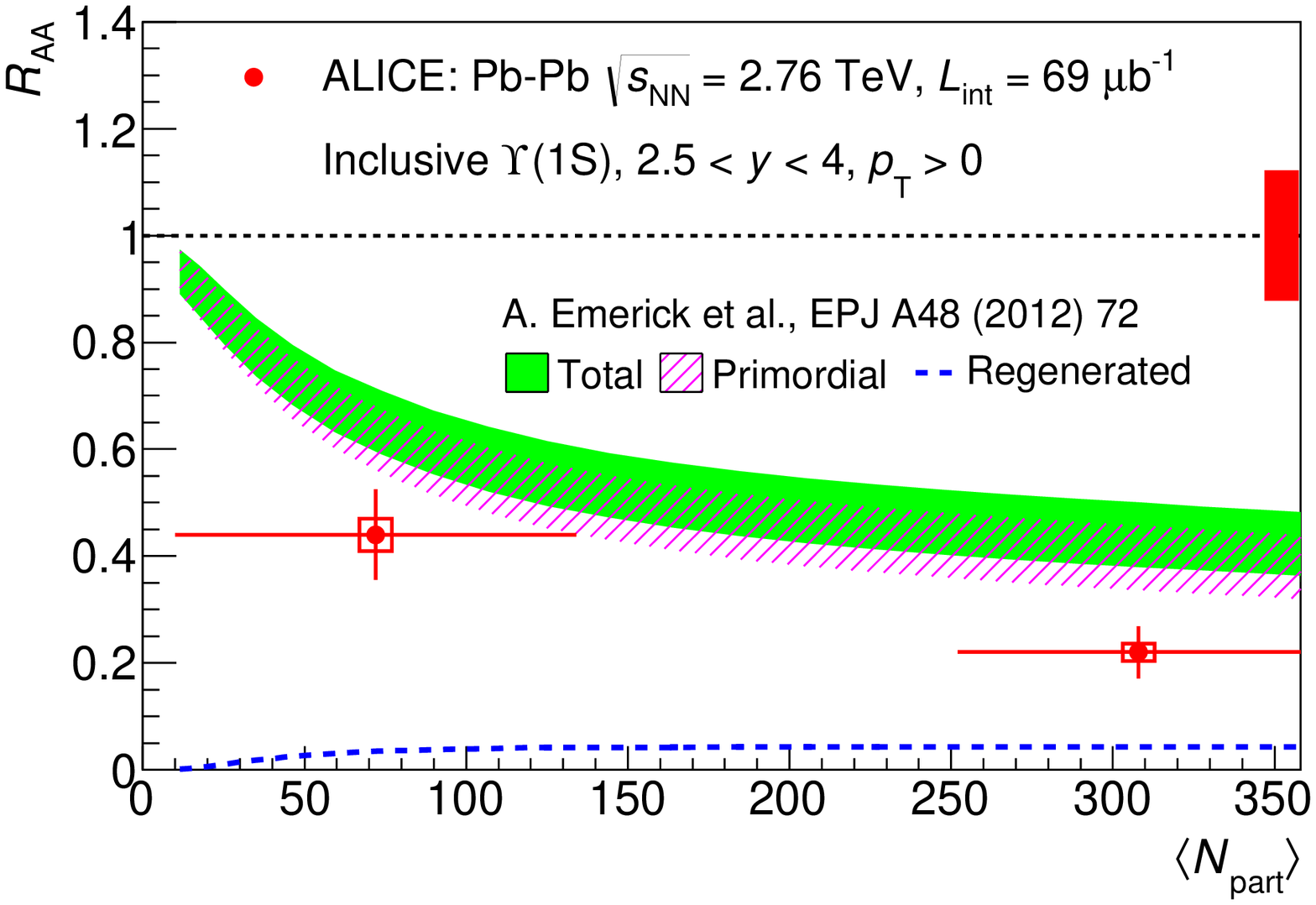}
\includegraphics[width=11cm]{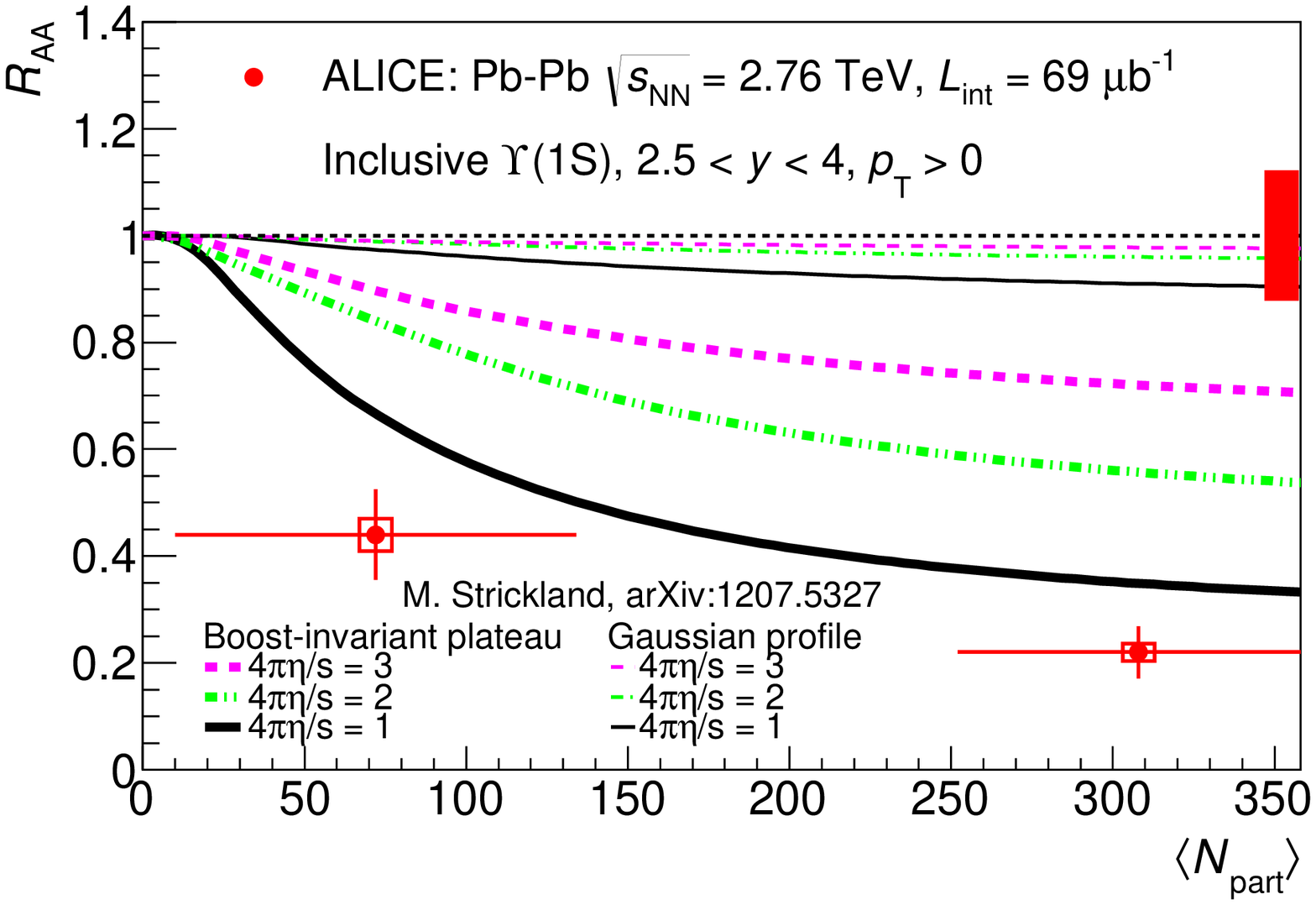}
\caption{Inclusive \state \raa as a function of $\langle N_{\rm part}\rangle$, compared with calculations from a transport~\cite{Emerick:2011xu} (top) and a dynamical~\cite{Strickland:2012cq} (bottom) model (see text for details). The same conventions as in Fig.~\ref{ALICEJpsiFig} are used to show the uncertainties.}
\label{ALICECentreFig}
\end{center}
\end{figure}

\begin{figure}[h*]
\begin{center}
\includegraphics[width=11cm]{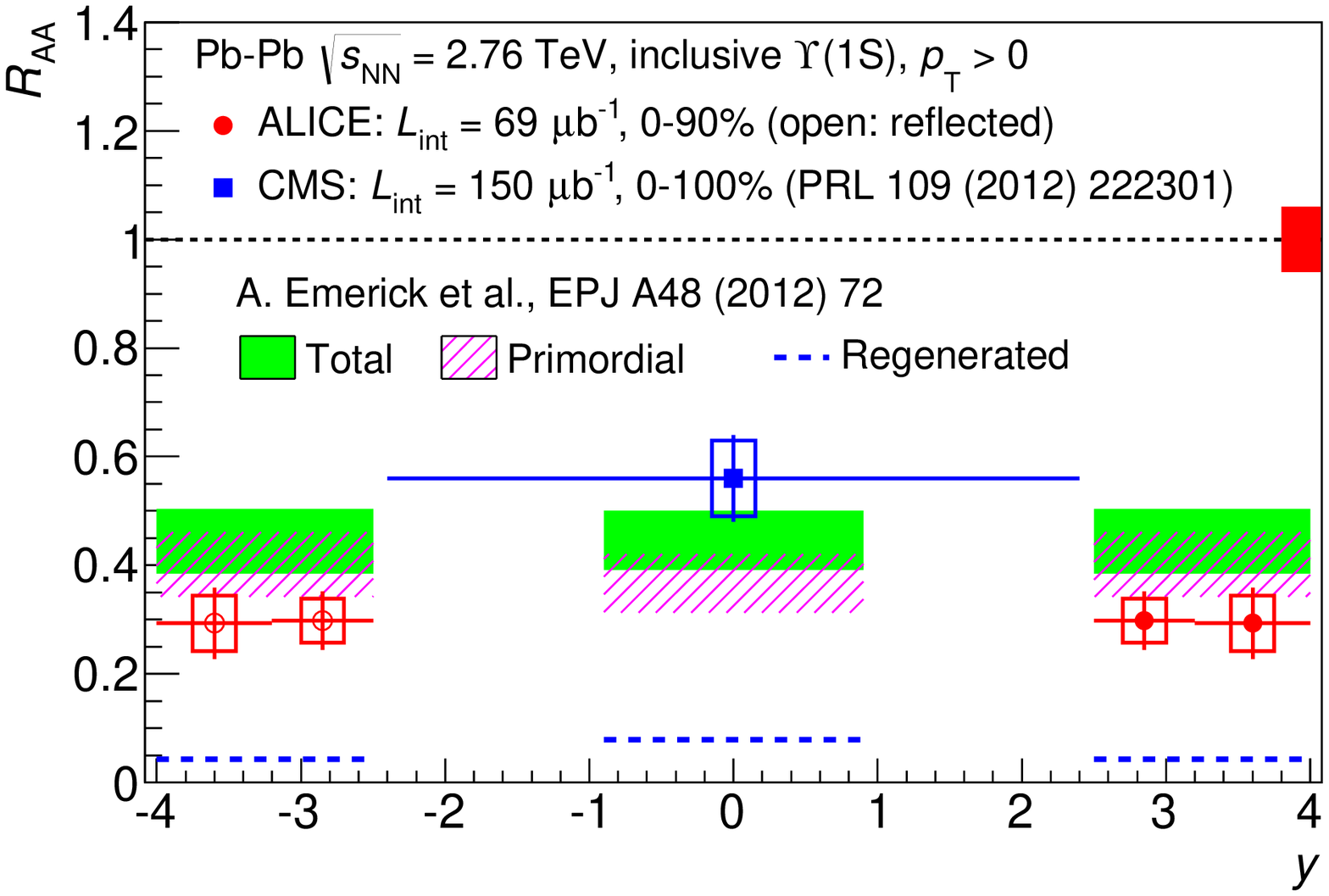}
\includegraphics[width=11cm]{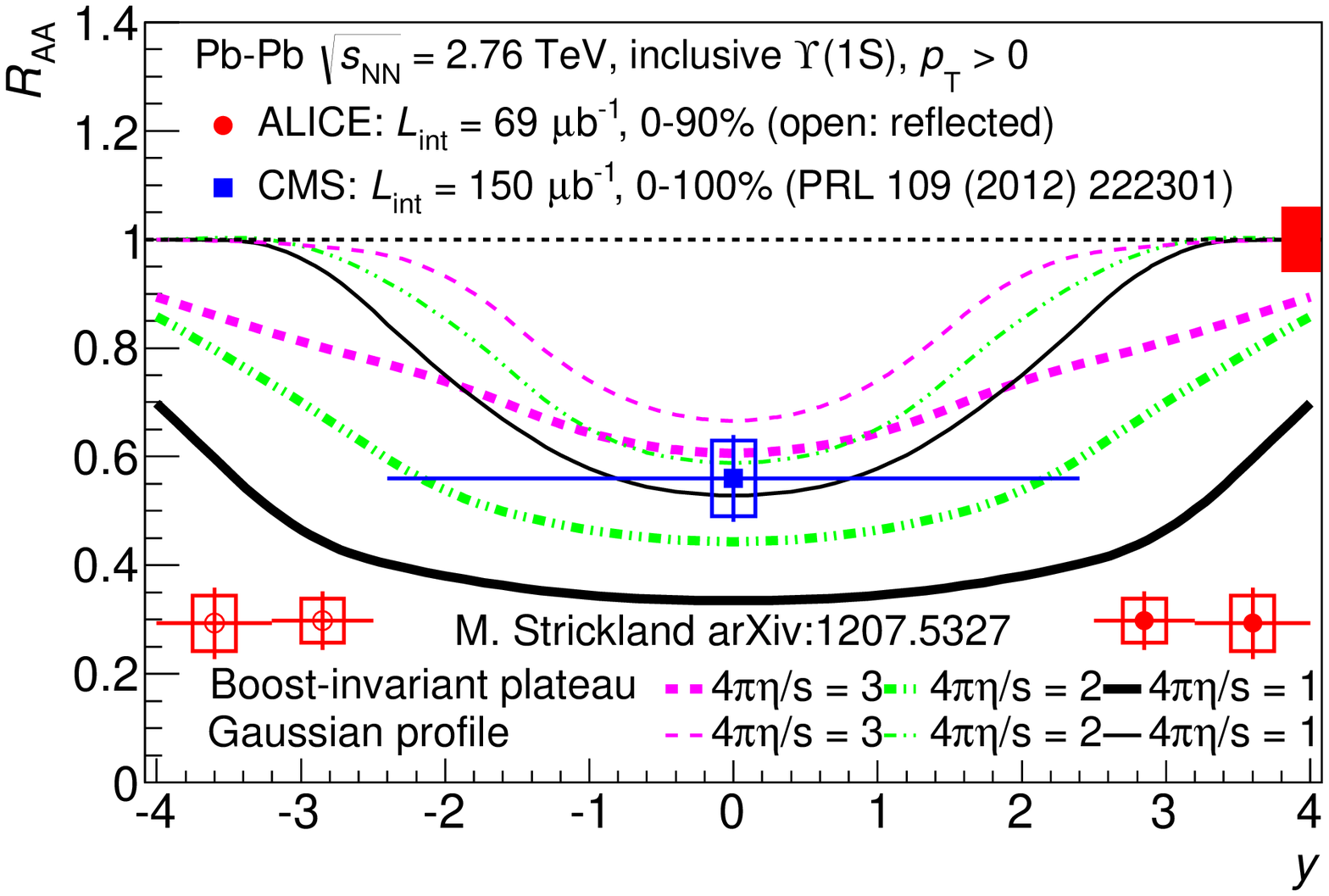}
\caption{Inclusive \state \raa as a function of rapidity measured in Pb\---Pb collisions at \sNN by ALICE in $2.5<y<4$ and CMS~\cite{Chatrchyan:2012lxa} in $|y|<2.4$, compared with the calculations from a transport~\cite{Emerick:2011xu,RefEmRap} (top) and a dynamical~\cite{Strickland:2012cq} (bottom) model (see text for details). Open points are reflected with respect to the measured ones and the same conventions as in Fig.~\ref{ALICEJpsiFig} are used to show the uncertainties. The relative correlated uncertainty on the ALICE measurement is $7\%$ (and is shown as a box at unity).}
\label{ALICECMSFig}
\end{center}
\end{figure}

\section{Results}

The $p_{\rm T}$-integrated nuclear modification factor measured in the rapidity interval $2.5<y<4$ is $0.30\pm0.05{\rm (stat)}\pm0.04{\rm (syst)}$ for the 0\---90\% centrality range and indicates a strong suppression of the inclusive $\Upsilon$(1S) production. The numerical values of the nuclear modification factor for the other centrality and rapidity intervals considered in the analysis are given in Table~\ref{ValueRAA}.

In Fig.~\ref{ALICEJpsiFig}, the \raa is shown as a function of $\langle N_{\rm part}\rangle$. Since our centrality intervals are large, a horizontal error bar was assigned point-to-point. It corresponds to the RMS of the $N_{\rm part}$ distribution~\cite{Abelev:2013qoq}. The observed suppression tends to be more pronounced in central (0\---20\%) than in semi-peripheral (20\---90\%) collisions. The CMS~\cite{Chatrchyan:2012lxa} data in $|y|<2.4$ 
are shown in the same figure. In central collisions, the suppression seems stronger at forward rapidity than at mid-rapidity. In semi-peripheral collisions, a similar effect might be present with a smaller significance.

In Fig.~\ref{ALICECentreFig}, the ALICE results are compared with the calculations from a transport~\cite{Emerick:2011xu,RefEmRap} (top) and a dynamical~\cite{Strickland:2012cq} (bottom) model. The transport model~\cite{Emerick:2011xu} employs a kinetic rate-equation approach in an evolving QGP and includes both suppression and regeneration effects. 
In the model~\cite{Emerick:2011xu}, CNM effects were calculated by varying an effective absorption cross section between $0$ and $2\,{\rm mb}$, resulting in an uncertainty band used to represent the $R_{\rm AA}$. The transport model clearly underestimates the observed suppression, even if the shape of the centrality dependence is fairly reproduced. The dynamical model~\cite{Strickland:2012cq} does not include CNM or regeneration effects. The calculation of the bottomonium suppression is based on a complex-potential approach in an evolving QGP described with a hydrodynamical model. 
It is assumed that the initial temperature profile in rapidity is a boost-invariant plateau, as inferred from the Bjorken picture~\cite{Bjorken:1982qr} of heavy-ion collisions. The results obtained with a Gaussian profile corresponding to the Landau picture~\cite{Landaucite} are also shown. Three values of plasma shear viscosity to entropy density ratio ($4\pi\eta{\rm /s}$) are used in the calculations, including the limiting case where $4\pi\eta{\rm /s}=1$. The model calculations
underestimate the measured suppression, independently of the temperature profiles and the model parameter assumptions adopted. The result calculated with $4\pi\eta{\rm /s}=1$ in the Bjorken scenario shows the largest suppression and fairly reproduces the shape of the data. It has to be noted that the comparison between the $R_{\rm AA}$ values and theoretical predictions depends on whether the results are shown as a function of $\langle N_{\rm part}\rangle$ or $\langle N_{\rm part}^{\rm w}\rangle$. In particular, if $\langle N_{\rm part}^{\rm w}\rangle$ is adopted, the semi-peripheral $R_{\rm AA}$ data point is fairly described by both the transport and the dynamical models.

The rapidity dependence of the inclusive \state $R_{\rm AA}$, integrated over centrality (0\---90\%) for $p_{\rm T}>0$, is presented in Fig.~\ref{ALICECMSFig}. The ALICE results are compared with those of CMS~\cite{Chatrchyan:2012lxa} ($|y|<2.4$). The observed suppression seems stronger at forward than at mid-rapidity. 

The predictions of the transport model~\cite{Emerick:2011xu,RefEmRap} are also shown in Fig.~\ref{ALICECMSFig} (top). The model predicts a nearly constant $R_{\rm AA}$ as a function of the rapidity which is in disagreement with CMS and ALICE data.
In Fig.~\ref{ALICECMSFig} (bottom), the data are compared with the calculations of the dynamical model~\cite{Strickland:2012cq}. All parameter sets used in the model calculations predict a rapidity dependence which is the opposite of the measured one.

In both the transport and the dynamical models, the inclusive \state suppression is largely due to the in-medium dissociation of higher mass bottomonia. The even larger suppression observed in the ALICE data might then point to a significant dissociation of direct $\Upsilon$(1S). However, to reach a more quantitative assessment, the role played by CNM effects at forward rapidity should be more accurately verified and constrained by data.

\section{Conclusions}

In summary, we have presented the measurement of the nuclear modification factor of inclusive \state production at forward rapidity ($2.5<y<4$) and down to zero transverse momentum ($p_{\rm T}>0$) in Pb\---Pb collisions at $\sqrt{s_{\rm NN}}=2.76\,{\rm TeV}$. The observed suppression of inclusive \state seems stronger in central (0\---20\%) than in semi-peripheral (20\---90\%) collisions and tends to show a pronounced rapidity dependence over the large domain covered by ALICE ($2.5<y<4$) and CMS ($|y|<2.4$). The ALICE inclusive \state suppression is underestimated by the transport model~\cite{Emerick:2011xu,RefEmRap} as well as by the dynamical model~\cite{Strickland:2012cq} considered in this Letter. The suppression predicted by the transport model calculations is approximately constant with rapidity while the measured one is more pronounced at forward than at mid-rapidity. In the case of the dynamical model, the calculated rapidity trend is the opposite of the observed one. A precise measurement of \state feed-down from higher mass 
bottomonia, as well as an accurate estimate of CNM effects in the kinematic range probed by ALICE is required in order to make a more stringent comparison with models. The \state production in p\---A collisions has recently been measured with the ALICE muon spectrometer~\cite{pPb1Spaper} and should help to gain further insight on the size of 
the CNM effects.               
\bibliographystyle{utphys} 
\bibliography{UpsiPbPb_bibli}

\providecommand{\href}[2]{#2}\begingroup\raggedright\begin{thebibliography}{10}

\bibitem{Shuryak:1980tp}
E.~V. Shuryak, ``{Quantum Chromodynamics and the Theory of Superdense
  Matter},''
\href{http://dx.doi.org/10.1016/0370-1573(80)90105-2}{{\em Phys.Rept.}
  {\bfseries 61} (1980) 71--158}.

\bibitem{Cheng:2009zi}
M.~Cheng, S.~Ejiri, P.~Hegde, F.~Karsch, O.~Kaczmarek, {\em et~al.},
  ``{Equation of State for physical quark masses},''
  \href{http://dx.doi.org/10.1103/PhysRevD.81.054504}{{\em Phys.Rev.}
  {\bfseries D81} (2010) 054504},
\href{http://arxiv.org/abs/0911.2215}{{\ttfamily arXiv:0911.2215 [hep-lat]}}.

\bibitem{Borsanyi:2010bp}
{\bfseries Wuppertal-Budapest} Collaboration, S.~Borsanyi {\em et~al.}, ``{Is
  there still any $T_c$ mystery in lattice QCD? Results with physical masses in
  the continuum limit III},''
  \href{http://dx.doi.org/10.1007/JHEP09(2010)073}{{\em JHEP} {\bfseries 1009}
  (2010) 073},
\href{http://arxiv.org/abs/1005.3508}{{\ttfamily arXiv:1005.3508 [hep-lat]}}.

\bibitem{Matsui:1986dk}
T.~Matsui and H.~Satz, ``{J/$\psi$ Suppression by Quark-Gluon Plasma
  Formation},''
\href{http://dx.doi.org/10.1016/0370-2693(86)91404-8}{{\em Phys.Lett.}
  {\bfseries B178} (1986) 416}.

\bibitem{Digal:2001ue}
S.~Digal, P.~Petreczky, and H.~Satz, ``{Quarkonium feed down and sequential
  suppression},'' \href{http://dx.doi.org/10.1103/PhysRevD.64.094015}{{\em
  Phys.Rev.} {\bfseries D64} (2001) 094015},
\href{http://arxiv.org/abs/hep-ph/0106017}{{\ttfamily arXiv:hep-ph/0106017
  [hep-ph]}}.

\bibitem{Wong:2004zr}
C.-Y. Wong, ``{Heavy quarkonia in quark-gluon plasma},''
  \href{http://dx.doi.org/10.1103/PhysRevC.72.034906}{{\em Phys.Rev.}
  {\bfseries C72} (2005) 034906},
\href{http://arxiv.org/abs/hep-ph/0408020}{{\ttfamily arXiv:hep-ph/0408020
  [hep-ph]}}.

\bibitem{Satz:2005hx}
H.~Satz, ``{Colour deconfinement and quarkonium binding},''
  \href{http://dx.doi.org/10.1088/0954-3899/32/3/R01}{{\em J.Phys.} {\bfseries
  G32} (2006) R25},
\href{http://arxiv.org/abs/hep-ph/0512217}{{\ttfamily arXiv:hep-ph/0512217
  [hep-ph]}}.

\bibitem{Mocsy:2007jz}
A.~Mocsy and P.~Petreczky, ``{Color screening melts quarkonium},''
  \href{http://dx.doi.org/10.1103/PhysRevLett.99.211602}{{\em Phys.Rev.Lett.}
  {\bfseries 99} (2007) 211602},
\href{http://arxiv.org/abs/0706.2183}{{\ttfamily arXiv:0706.2183 [hep-ph]}}.

\bibitem{Petreczky:2010tk}
P.~Petreczky, C.~Miao, and A.~Mocsy, ``{Quarkonium spectral functions with
  complex potential},'' {\em Nucl.Phys.} {\bfseries A855} .

\bibitem{Abreu:1998wx}
{\bfseries NA38} Collaboration, M.~Abreu {\em et~al.}, ``{J /$\psi$, $\psi$'
  and Drell-Yan production in S U interactions at 200-GeV per nucleon},''
\href{http://dx.doi.org/10.1016/S0370-2693(99)00057-X}{{\em Phys.Lett.}
  {\bfseries B449} (1999) 128--136}.

\bibitem{Alessandro:2004ap}
{\bfseries NA50} Collaboration, B.~Alessandro {\em et~al.}, ``{A New
  measurement of J/$\psi$ suppression in Pb-Pb collisions at 158-GeV per
  nucleon},'' \href{http://dx.doi.org/10.1140/epjc/s2004-02107-9}{{\em
  Eur.Phys.J.} {\bfseries C39} (2005) 335--345},
\href{http://arxiv.org/abs/hep-ex/0412036}{{\ttfamily arXiv:hep-ex/0412036
  [hep-ex]}}.

\bibitem{Arnaldi:2007zz}
{\bfseries NA60} Collaboration, R.~Arnaldi {\em et~al.}, ``{J/$\psi$ production
  in indium-indium collisions at 158-GeV/nucleon},''
\href{http://dx.doi.org/10.1103/PhysRevLett.99.132302}{{\em Phys.Rev.Lett.}
  {\bfseries 99} (2007) 132302}.

\bibitem{Adare:2006ns}
{\bfseries PHENIX} Collaboration, A.~Adare {\em et~al.}, ``{J/$\psi$ Production
  vs Centrality, Transverse Momentum, and Rapidity in Au+Au Collisions at
  $\sqrt{s_{NN}} = 200$ GeV},''
  \href{http://dx.doi.org/10.1103/PhysRevLett.98.232301}{{\em Phys.Rev.Lett.}
  {\bfseries 98} (2007) 232301},
\href{http://arxiv.org/abs/nucl-ex/0611020}{{\ttfamily arXiv:nucl-ex/0611020
  [nucl-ex]}}.

\bibitem{Abelev:2009qaa}
{\bfseries STAR} Collaboration, B.~Abelev {\em et~al.}, ``{J/$\psi$ production
  at high transverse momentum in p+p and Cu+Cu collisions at $\sqrt{s_{NN}}$ =
  200GeV},'' \href{http://dx.doi.org/10.1103/PhysRevC.80.041902}{{\em
  Phys.Rev.} {\bfseries C80} (2009) 041902},
\href{http://arxiv.org/abs/0904.0439}{{\ttfamily arXiv:0904.0439 [nucl-ex]}}.

\bibitem{Abelev:2012rv}
{\bfseries ALICE} Collaboration, B.~Abelev {\em et~al.}, ``{J/$\psi$
  suppression at forward rapidity in Pb-Pb collisions at $\sqrt{s_{NN}}=2.76$
  TeV},'' \href{http://dx.doi.org/10.1103/PhysRevLett.109.072301}{{\em
  Phys.Rev.Lett.} {\bfseries 109} (2012) 072301},
\href{http://arxiv.org/abs/1202.1383}{{\ttfamily arXiv:1202.1383 [hep-ex]}}.

\bibitem{Aad:2010aa}
{\bfseries ATLAS} Collaboration, G.~Aad {\em et~al.}, ``{Measurement of the
  centrality dependence of J$/\psi$ yields and observation of Z production in
  lead-lead collisions with the ATLAS detector at the LHC},''
  \href{http://dx.doi.org/10.1016/j.physletb.2011.02.006}{{\em Phys.Lett.}
  {\bfseries B697} (2011) 294--312},
\href{http://arxiv.org/abs/1012.5419}{{\ttfamily arXiv:1012.5419 [hep-ex]}}.

\bibitem{Chatrchyan:2012np}
{\bfseries CMS} Collaboration, S.~Chatrchyan {\em et~al.}, ``{Suppression of
  non-prompt J/$\psi$, prompt J/$\psi$, and $\Upsilon$(1S) in PbPb collisions
  at $\sqrt{s_{NN}}=2.76$ TeV},''
  \href{http://dx.doi.org/10.1007/JHEP05(2012)063}{{\em JHEP} {\bfseries 1205}
  (2012) 063},
\href{http://arxiv.org/abs/1201.5069}{{\ttfamily arXiv:1201.5069 [nucl-ex]}}.

\bibitem{Grandchamp:2003uw}
L.~Grandchamp, R.~Rapp, and G.~E. Brown, ``{In medium effects on charmonium
  production in heavy ion collisions},''
  \href{http://dx.doi.org/10.1103/PhysRevLett.92.212301}{{\em Phys.Rev.Lett.}
  {\bfseries 92} (2004) 212301},
\href{http://arxiv.org/abs/hep-ph/0306077}{{\ttfamily arXiv:hep-ph/0306077
  [hep-ph]}}.

\bibitem{Bratkovskaya:2004cq}
E.~Bratkovskaya, A.~Kostyuk, W.~Cassing, and H.~Stoecker, ``{Charmonium
  chemistry in A+A collisions at relativistic energies},''
  \href{http://dx.doi.org/10.1103/PhysRevC.69.054903}{{\em Phys.Rev.}
  {\bfseries C69} (2004) 054903},
\href{http://arxiv.org/abs/nucl-th/0402042}{{\ttfamily arXiv:nucl-th/0402042
  [nucl-th]}}.

\bibitem{Thews:2000rj}
R.~L. Thews, M.~Schroedter, and J.~Rafelski, ``{Enhanced J/$\psi$ production in
  deconfined quark matter},''
  \href{http://dx.doi.org/10.1103/PhysRevC.63.054905}{{\em Phys.Rev.}
  {\bfseries C63} (2001) 054905},
\href{http://arxiv.org/abs/hep-ph/0007323}{{\ttfamily arXiv:hep-ph/0007323
  [hep-ph]}}.

\bibitem{BraunMunzinger:2000px}
P.~Braun-Munzinger and J.~Stachel, ``{(Non)thermal aspects of charmonium
  production and a new look at J/$\psi$ suppression},''
  \href{http://dx.doi.org/10.1016/S0370-2693(00)00991-6}{{\em Phys.Lett.}
  {\bfseries B490} (2000) 196--202},
\href{http://arxiv.org/abs/nucl-th/0007059}{{\ttfamily arXiv:nucl-th/0007059
  [nucl-th]}}.

\bibitem{Andronic:2003zv}
A.~Andronic, P.~Braun-Munzinger, K.~Redlich, and J.~Stachel, ``{Statistical
  hadronization of charm in heavy ion collisions at SPS, RHIC and LHC},''
  \href{http://dx.doi.org/10.1016/j.physletb.2003.07.066}{{\em Phys.Lett.}
  {\bfseries B571} (2003) 36--44},
\href{http://arxiv.org/abs/nucl-th/0303036}{{\ttfamily arXiv:nucl-th/0303036
  [nucl-th]}}.

\bibitem{Alessandro:2006ju}
{\bfseries NA50} Collaboration, B.~Alessandro {\em et~al.}, ``{$\psi$'
  production in Pb-Pb collisions at 158-GeV/nucleon},''
  \href{http://dx.doi.org/10.1140/epjc/s10052-006-0153-y}{{\em Eur.Phys.J.}
  {\bfseries C49} (2007) 559--567},
\href{http://arxiv.org/abs/nucl-ex/0612013}{{\ttfamily arXiv:nucl-ex/0612013
  [nucl-ex]}}.

\bibitem{Adamczyk:2013poh}
{\bfseries STAR} Collaboration, L.~Adamczyk {\em et~al.}, ``{Suppression of
  $\Upsilon$ Production in d+Au and Au+Au Collisions at $\sqrt{s_{NN}}$ = 200
  GeV},'' \href{http://dx.doi.org/10.1016/j.physletb.2014.06.028}{{\em
  Phys.Lett.} {\bfseries B735} (2014) 127},
\href{http://arxiv.org/abs/1312.3675}{{\ttfamily arXiv:1312.3675 [nucl-ex]}}.

\bibitem{PHENIXUps}
{\bfseries PHENIX} Collaboration, A.~Adare {\em et~al.}, ``{Measurement of
  $\Upsilon$(1S+2S+3S) production in $p$$+$$p$ and Au$+$Au collisions at
  $\sqrt{s_{NN}}=200$ GeV},''
\href{http://arxiv.org/abs/1404.2246}{{\ttfamily arXiv:1404.2246 [nucl-ex]}}.

\bibitem{Chatrchyan:2011pe}
{\bfseries CMS} Collaboration, S.~Chatrchyan {\em et~al.}, ``{Indications of
  suppression of excited $\Upsilon$ states in PbPb collisions at
  $\sqrt{s_{NN}}$ = 2.76 TeV},''
  \href{http://dx.doi.org/10.1103/PhysRevLett.107.052302}{{\em Phys.Rev.Lett.}
  {\bfseries 107} (2011) 052302},
\href{http://arxiv.org/abs/1105.4894}{{\ttfamily arXiv:1105.4894 [nucl-ex]}}.

\bibitem{Chatrchyan:2012lxa}
{\bfseries CMS} Collaboration, S.~Chatrchyan {\em et~al.}, ``{Observation of
  sequential $\Upsilon$ suppression in PbPb collisions},''
  \href{http://dx.doi.org/10.1103/PhysRevLett.109.222301}{{\em Phys.Rev.Lett.}
  {\bfseries 109} (2012) 222301},
\href{http://arxiv.org/abs/1208.2826}{{\ttfamily arXiv:1208.2826 [nucl-ex]}}.

\bibitem{PN}
{\bfseries ALICE} Collaboration, B.~Abelev {\em et~al.}, ``{Production of
  $\Upsilon$(1S) in PbPb collisions at $\sqrt{s_{\rm NN}}=2.76$ TeV},'' {\em
  \href{https://cds.cern.ch/record/1727966/files/ALICE-PN-2014-001.pdf}{ALICE-PUBLIC-2014-001}}
  .

\bibitem{RefEmRap}
L.~Grandchamp, S.~Lumpkins, D.~Sun, H.~van Hees, and R.~Rapp, ``{Bottomonium
  production at RHIC and CERN LHC},''
  \href{http://dx.doi.org/10.1103/PhysRevC.73.064906}{{\em Phys.Rev.}
  {\bfseries C73} (2006) 064906},
\href{http://arxiv.org/abs/hep-ph/0507314}{{\ttfamily arXiv:hep-ph/0507314
  [hep-ph]}}.

\bibitem{QuarkoppALICE}
{\bfseries ALICE} Collaboration, B.~B. Abelev {\em et~al.}, ``{Measurement of
  quarkonium production at forward rapidity in pp collisions at $\sqrt{s}$= 7
  TeV},'' {\em Eur.Phys.J.} {\bfseries C74} (2014) 2974,
\href{http://arxiv.org/abs/1403.3648}{{\ttfamily arXiv:1403.3648 [nucl-ex]}}.

\bibitem{Emerick:2011xu}
A.~Emerick, X.~Zhao, and R.~Rapp, ``{Bottomonia in the Quark-Gluon Plasma and
  their Production at RHIC and LHC},''
  \href{http://dx.doi.org/10.1140/epja/i2012-12072-y}{{\em Eur.Phys.J.}
  {\bfseries A48} (2012) 72},
\href{http://arxiv.org/abs/1111.6537}{{\ttfamily arXiv:1111.6537 [hep-ph]}}.

\bibitem{Strickland:2012cq}
M.~Strickland, ``{Thermal Bottomonium Suppression},''
  \href{http://dx.doi.org/10.1063/1.4795953}{{\em AIP Conf.Proc.} {\bfseries
  1520} (2013) 179--184},
\href{http://arxiv.org/abs/1207.5327}{{\ttfamily arXiv:1207.5327 [hep-ph]}}.

\bibitem{Aamodt:2008zz}
{\bfseries ALICE} Collaboration, K.~Aamodt {\em et~al.}, ``{The ALICE
  experiment at the CERN LHC},''
\href{http://dx.doi.org/10.1088/1748-0221/3/08/S08002}{{\em JINST} {\bfseries
  3} (2008) S08002}.

\bibitem{VZEROJINST}
{\bfseries ALICE} Collaboration, E.~Abbas {\em et~al.}, ``{Performance of the
  ALICE VZERO system},''
  \href{http://dx.doi.org/10.1088/1748-0221/8/10/P10016}{{\em JINST} {\bfseries
  8} (2013) P10016},
\href{http://arxiv.org/abs/1306.3130}{{\ttfamily arXiv:1306.3130 [nucl-ex]}}.

\bibitem{ALICEPerfPaper}
{\bfseries ALICE} Collaboration, B.~B. Abelev {\em et~al.}, ``{Performance of
  the ALICE Experiment at the CERN LHC},'' {\em Int.J.Mod.Phys.} {\bfseries
  A29} (2014) 1430044,
\href{http://arxiv.org/abs/1402.4476}{{\ttfamily arXiv:1402.4476 [nucl-ex]}}.

\bibitem{Abelev:2013qoq}
{\bfseries ALICE} Collaboration, B.~Abelev {\em et~al.}, ``{Centrality
  determination of Pb-Pb collisions at $\sqrt{s_{NN}}$ = 2.76 TeV with
  ALICE},'' \href{http://dx.doi.org/10.1103/PhysRevC.88.044909}{{\em Phys.Rev.}
  {\bfseries C88} (2013) 044909},
\href{http://arxiv.org/abs/1301.4361}{{\ttfamily arXiv:1301.4361 [nucl-ex]}}.

\bibitem{Aamodt:2010cz}
{\bfseries ALICE} Collaboration, K.~Aamodt {\em et~al.}, ``{Centrality
  dependence of the charged-particle multiplicity density at mid-rapidity in
  Pb-Pb collisions at $\sqrt{s_{NN}}=2.76$ TeV},''
  \href{http://dx.doi.org/10.1103/PhysRevLett.106.032301}{{\em Phys.Rev.Lett.}
  {\bfseries 106} (2011) 032301},
\href{http://arxiv.org/abs/1012.1657}{{\ttfamily arXiv:1012.1657 [nucl-ex]}}.

\bibitem{Abelev:2013ila}
{\bfseries ALICE} Collaboration, B.~B. Abelev {\em et~al.}, ``{Centrality,
  rapidity and transverse momentum dependence of J/$\psi$ suppression in Pb-Pb
  collisions at $\sqrt{s_{NN}}$=2.76 TeV},''
  \href{http://dx.doi.org/10.1016/j.physletb.2014.05.064}{{\em Phys.Lett.}
  {\bfseries 743} (2014) 314--327},
\href{http://arxiv.org/abs/1311.0214}{{\ttfamily arXiv:1311.0214 [nucl-ex]}}.

\bibitem{Aamodt:2011gj}
{\bfseries ALICE} Collaboration, K.~Aamodt {\em et~al.}, ``{Rapidity and
  transverse momentum dependence of inclusive J/$\psi$ production in $pp$
  collisions at $\sqrt{s} = 7$ TeV},''
  \href{http://dx.doi.org/10.1016/j.physletb.2011.09.054,
  10.1016/j.physletb.2012.10.060}{{\em Phys.Lett.} {\bfseries B704} (2011)
  442--455},
\href{http://arxiv.org/abs/1105.0380}{{\ttfamily arXiv:1105.0380 [hep-ex]}}.

\bibitem{CB2}
J.~Gaiser {\em SLAC Stanford - SLAC-255 (82.REC.JUN.83) 194p,
  http://www/slac.stanford.edu/cgi-wrap/getdoc/slac-r-255.pdf} .

\bibitem{PDGTabRef}
{\bfseries Particle Data Group} Collaboration, J.~Beringer {\em et~al.},
  ``{Review of Particle Physics (RPP)},''
\href{http://dx.doi.org/10.1103/PhysRevD.86.010001}{{\em Phys.Rev.} {\bfseries
  D86} (2012) 010001}.

\bibitem{Acosta:2001gv}
{\bfseries CDF} Collaboration, D.~Acosta {\em et~al.}, ``{$\Upsilon$ production
  and polarization in $p\bar{p}$ collisions at $\sqrt{s}=$ 1.8-TeV},''
\href{http://dx.doi.org/10.1103/PhysRevLett.88.161802}{{\em Phys.Rev.Lett.}
  {\bfseries 88} (2002) 161802}.

\bibitem{LHCb:2012aa}
{\bfseries LHCb} Collaboration, R.~Aaij {\em et~al.}, ``{Measurement of
  $\Upsilon$ production in pp collisions at $\sqrt{s}$ = 7 TeV},''
  \href{http://dx.doi.org/10.1140/epjc/s10052-012-2025-y}{{\em Eur.Phys.J.}
  {\bfseries C72} (2012) 2025},
\href{http://arxiv.org/abs/1202.6579}{{\ttfamily arXiv:1202.6579 [hep-ex]}}.

\bibitem{Khachatryan:2010zg}
{\bfseries CMS} Collaboration, V.~Khachatryan {\em et~al.}, ``{Measurement of
  the Inclusive $\Upsilon$ production cross section in $pp$ collisions at
  $\sqrt{s}=7$ TeV},'' \href{http://dx.doi.org/10.1103/PhysRevD.83.112004}{{\em
  Phys.Rev.} {\bfseries D83} (2011) 112004},
\href{http://arxiv.org/abs/1012.5545}{{\ttfamily arXiv:1012.5545 [hep-ex]}}.

\bibitem{ParamUpSystRef}
F.~Bossu, Z.~C. del Valle, A.~de~Falco, M.~Gagliardi, S.~Grigoryan, {\em
  et~al.}, ``{Phenomenological interpolation of the inclusive J/$\psi$ cross
  section to proton-proton collisions at 2.76 TeV and 5.5 TeV},''
\href{http://arxiv.org/abs/1103.2394}{{\ttfamily arXiv:1103.2394 [nucl-ex]}}.

\bibitem{Eskola:1998df}
K.~Eskola, V.~Kolhinen, and C.~Salgado, ``{The Scale dependent nuclear effects
  in parton distributions for practical applications},''
  \href{http://dx.doi.org/10.1007/s100520050513}{{\em Eur.Phys.J.} {\bfseries
  C9} (1999) 61--68},
\href{http://arxiv.org/abs/hep-ph/9807297}{{\ttfamily arXiv:hep-ph/9807297
  [hep-ph]}}.

\bibitem{Abazov:2008aa}
{\bfseries D0} Collaboration, V.~Abazov {\em et~al.}, ``{Measurement of the
  polarization of the $\Upsilon$(1S) and $\Upsilon$(2S) states in $p \bar{p}$
  collisions at $\sqrt{s}$ = 1.96-TeV},''
  \href{http://dx.doi.org/10.1103/PhysRevLett.101.182004}{{\em Phys.Rev.Lett.}
  {\bfseries 101} (2008) 182004},
\href{http://arxiv.org/abs/0804.2799}{{\ttfamily arXiv:0804.2799 [hep-ex]}}.

\bibitem{CDF:2011ag}
{\bfseries CDF} Collaboration, T.~Aaltonen {\em et~al.}, ``{Measurements of
  Angular Distributions of Muons From $\Upsilon$ Meson Decays in $p\bar{p}$
  Collisions at $\sqrt{s}=1.96$ TeV},''
  \href{http://dx.doi.org/10.1103/PhysRevLett.108.151802}{{\em Phys.Rev.Lett.}
  {\bfseries 108} (2012) 151802},
\href{http://arxiv.org/abs/1112.1591}{{\ttfamily arXiv:1112.1591 [hep-ex]}}.

\bibitem{Chatrchyan:2012woa}
{\bfseries CMS} Collaboration, S.~Chatrchyan {\em et~al.}, ``{Measurement of
  the $\Upsilon$(1S), $\Upsilon$(2S) and $\Upsilon$(3S) polarizations in $pp$
  collisions at $\sqrt{s}=7$ TeV},''
  \href{http://dx.doi.org/10.1103/PhysRevLett.110.081802}{{\em Phys.Rev.Lett.}
  {\bfseries 110} (2013) 081802},
\href{http://arxiv.org/abs/1209.2922}{{\ttfamily arXiv:1209.2922 [hep-ex]}}.

\bibitem{Aaij:2014nwa}
{\bfseries LHCb collaboration} Collaboration, R.~Aaij {\em et~al.},
  ``{Measurement of $\Upsilon$ production in $pp$ collisions at $\sqrt{s}=2.76$
  TeV},'' \href{http://dx.doi.org/10.1140/epjc/s10052-014-2835-1}{{\em
  Eur.Phys.J.} {\bfseries C74} (2014) 2835},
\href{http://arxiv.org/abs/1402.2539}{{\ttfamily arXiv:1402.2539 [hep-ex]}}.

\bibitem{Bjorken:1982qr}
J.~Bjorken, ``{Highly Relativistic Nucleus-Nucleus Collisions: The Central
  Rapidity Region},''
\href{http://dx.doi.org/10.1103/PhysRevD.27.140}{{\em Phys.Rev.} {\bfseries
  D27} (1983) 140--151}.

\bibitem{Landaucite}
L.~D. Landau, ``{On the multiple production of particles in high energy
  collisions},'' {\em Izv. Akad. Nauk SSSR, Ser. Fiz.} {\bfseries 17} (1953)
  51--64.

\bibitem{pPb1Spaper}
{\bfseries ALICE} Collaboration, B.~Abelev {\em et~al.}, ``{Production of
  inclusive $\Upsilon$(1S) and $\Upsilon$(2S) in p-Pb collisions at
  $\sqrt(s_{NN})$ = 5.02 TeV, CERN-PH-EP-2014-196},''.

\end{thebibliography}\endgroup
%
\newenvironment{acknowledgement}{\relax}{\relax}
\begin{acknowledgement}
\section*{Acknowledgements}
The ALICE collaboration would like to thank all its engineers and technicians for their invaluable contributions to the construction of the experiment and the CERN accelerator teams for the outstanding performance of the LHC complex.
The ALICE collaboration acknowledges the following funding agencies for their support in building and
running the ALICE detector:
State Committee of Science,  World Federation of Scientists (WFS)
and Swiss Fonds Kidagan, Armenia,
Conselho Nacional de Desenvolvimento Cient\'{\i}fico e Tecnol\'{o}gico (CNPq), Financiadora de Estudos e Projetos (FINEP),
Funda\c{c}\~{a}o de Amparo \`{a} Pesquisa do Estado de S\~{a}o Paulo (FAPESP);
National Natural Science Foundation of China (NSFC), the Chinese Ministry of Education (CMOE)
and the Ministry of Science and Technology of China (MSTC);
Ministry of Education and Youth of the Czech Republic;
Danish Natural Science Research Council, the Carlsberg Foundation and the Danish National Research Foundation;
The European Research Council under the European Community's Seventh Framework Programme;
Helsinki Institute of Physics and the Academy of Finland;
French CNRS-IN2P3, the `Region Pays de Loire', `Region Alsace', `Region Auvergne' and CEA, France;
German BMBF and the Helmholtz Association;
General Secretariat for Research and Technology, Ministry of
Development, Greece;
Hungarian OTKA and National Office for Research and Technology (NKTH);
Department of Atomic Energy and Department of Science and Technology of the Government of India;
Istituto Nazionale di Fisica Nucleare (INFN) and Centro Fermi -
Museo Storico della Fisica e Centro Studi e Ricerche "Enrico
Fermi", Italy;
MEXT Grant-in-Aid for Specially Promoted Research, Ja\-pan;
Joint Institute for Nuclear Research, Dubna;
National Research Foundation of Korea (NRF);
CONACYT, DGAPA, M\'{e}xico, ALFA-EC and the EPLANET Program
(European Particle Physics Latin American Network)
Stichting voor Fundamenteel Onderzoek der Materie (FOM) and the Nederlandse Organisatie voor Wetenschappelijk Onderzoek (NWO), Netherlands;
Research Council of Norway (NFR);
Polish Ministry of Science and Higher Education;
National Authority for Scientific Research - NASR (Autoritatea Na\c{t}ional\u{a} pentru Cercetare \c{S}tiin\c{t}ific\u{a} - ANCS);
Ministry of Education and Science of Russian Federation, Russian
Academy of Sciences, Russian Federal Agency of Atomic Energy,
Russian Federal Agency for Science and Innovations and The Russian
Foundation for Basic Research;
Ministry of Education of Slovakia;
Department of Science and Technology, South Africa;
CIEMAT, EELA, Ministerio de Econom\'{i}a y Competitividad (MINECO) of Spain, Xunta de Galicia (Conseller\'{\i}a de Educaci\'{o}n),
CEA\-DEN, Cubaenerg\'{\i}a, Cuba, and IAEA (International Atomic Energy Agency);
Swedish Research Council (VR) and Knut $\&$ Alice Wallenberg
Foundation (KAW);
Ukraine Ministry of Education and Science;
United Kingdom Science and Technology Facilities Council (STFC);
The United States Department of Energy, the United States National
Science Foundation, the State of Texas, and the State of Ohio.
\end{acknowledgement}
\appendix
\newpage
\section{The ALICE Collaboration}
\label{app:collab}



\begingroup
\small
\begin{flushleft}
B.~Abelev\Irefn{org71}\And
J.~Adam\Irefn{org37}\And
D.~Adamov\'{a}\Irefn{org79}\And
M.M.~Aggarwal\Irefn{org83}\And
M.~Agnello\Irefn{org90}\textsuperscript{,}\Irefn{org107}\And
A.~Agostinelli\Irefn{org26}\And
N.~Agrawal\Irefn{org44}\And
Z.~Ahammed\Irefn{org126}\And
N.~Ahmad\Irefn{org18}\And
I.~Ahmed\Irefn{org15}
% ...
\renewcommand\labelenumi{\textsuperscript{\theenumi}~}

\section*{Affiliation notes}
\renewcommand\theenumi{\roman{enumi}}
\begin{Authlist}
\item \Adef{0}Deceased
\item \Adef{idp6963648}{Also at: St. Petersburg State Polytechnical University}
\item \Adef{idp8811504}{Also at: Department of Applied Physics, Aligarh Muslim University, Aligarh, India}
\item \Adef{idp9492976}{Also at: M.V. Lomonosov Moscow State University, D.V. Skobeltsyn Institute of Nuclear Physics, Moscow, Russia}
\item \Adef{idp9744352}{Also at: University of Belgrade, Faculty of Physics and "Vin\v{c}a" Institute of Nuclear Sciences, Belgrade, Serbia}
\item \Adef{idp10049904}{Permanent Address: Permanent Address: Konkuk University, Seoul, Korea}
\item \Adef{idp10594192}{Also at: Institute of Theoretical Physics, University of Wroclaw, Wroclaw, Poland}
\item \Adef{idp11497024}{Also at: University of Kansas, Lawrence, KS, United States}
\end{Authlist}

\section*{Collaboration Institutes}
\renewcommand\theenumi{\arabic{enumi}~}
\begin{Authlist}

\item \Idef{org1}A.I. Alikhanyan National Science Laboratory (Yerevan Physics Institute) Foundation, Yerevan, Armenia
\item \Idef{org2}Benem\'{e}rita Universidad Aut\'{o}noma de Puebla, Puebla, Mexico
\item \Idef{org3}Bogolyubov Institute for Theoretical Physics, Kiev, Ukraine
\item \Idef{org4}Bose Institute, Department of Physics and Centre for Astroparticle Physics and Space Science (CAPSS), Kolkata, India
\item \Idef{org5}Budker Institute for Nuclear Physics, Novosibirsk, Russia
\item \Idef{org6}California Polytechnic State University, San Luis Obispo, CA, United States
\item \Idef{org7}Central China Normal University, Wuhan, China
\item \Idef{org8}Centre de Calcul de l'IN2P3, Villeurbanne, France
\item \Idef{org9}Centro de Aplicaciones Tecnol\'{o}gicas y Desarrollo Nuclear (CEADEN), Havana, Cuba
\item \Idef{org10}Centro de Investigaciones Energ\'{e}ticas Medioambientales y Tecnol\'{o}gicas (CIEMAT), Madrid, Spain
\item \Idef{org11}Centro de Investigaci\'{o}n y de Estudios Avanzados (CINVESTAV), Mexico City and M\'{e}rida, Mexico
\item \Idef{org12}Centro Fermi - Museo Storico della Fisica e Centro Studi e Ricerche ``Enrico Fermi'', Rome, Italy
\item \Idef{org13}Chicago State University, Chicago, USA
\item \Idef{org14}Commissariat \`{a} l'Energie Atomique, IRFU, Saclay, France
\item \Idef{org15}COMSATS Institute of Information Technology (CIIT), Islamabad, Pakistan
\item \Idef{org16}Departamento de F\'{\i}sica de Part\'{\i}culas and IGFAE, Universidad de Santiago de Compostela, Santiago de Compostela, Spain
\item \Idef{org17}Department of Physics and Technology, University of Bergen, Bergen, Norway
\item \Idef{org18}Department of Physics, Aligarh Muslim University, Aligarh, India
\item \Idef{org19}Department of Physics, Ohio State University, Columbus, OH, United States
\item \Idef{org20}Department of Physics, Sejong University, Seoul, South Korea
\item \Idef{org21}Department of Physics, University of Oslo, Oslo, Norway
\item \Idef{org22}Dipartimento di Fisica dell'Universit\`{a} 'La Sapienza' and Sezione INFN Rome, Italy
\item \Idef{org23}Dipartimento di Fisica dell'Universit\`{a} and Sezione INFN, Cagliari, Italy
\item \Idef{org24}Dipartimento di Fisica dell'Universit\`{a} and Sezione INFN, Trieste, Italy
\item \Idef{org25}Dipartimento di Fisica dell'Universit\`{a} and Sezione INFN, Turin, Italy
\item \Idef{org26}Dipartimento di Fisica e Astronomia dell'Universit\`{a} and Sezione INFN, Bologna, Italy
\item \Idef{org27}Dipartimento di Fisica e Astronomia dell'Universit\`{a} and Sezione INFN, Catania, Italy
\item \Idef{org28}Dipartimento di Fisica e Astronomia dell'Universit\`{a} and Sezione INFN, Padova, Italy
\item \Idef{org29}Dipartimento di Fisica `E.R.~Caianiello' dell'Universit\`{a} and Gruppo Collegato INFN, Salerno, Italy
\item \Idef{org30}Dipartimento di Scienze e Innovazione Tecnologica dell'Universit\`{a} del  Piemonte Orientale and Gruppo Collegato INFN, Alessandria, Italy
\item \Idef{org31}Dipartimento Interateneo di Fisica `M.~Merlin' and Sezione INFN, Bari, Italy
\item \Idef{org32}Division of Experimental High Energy Physics, University of Lund, Lund, Sweden
\item \Idef{org33}Eberhard Karls Universit\"{a}t T\"{u}bingen, T\"{u}bingen, Germany
\item \Idef{org34}European Organization for Nuclear Research (CERN), Geneva, Switzerland
\item \Idef{org35}Faculty of Engineering, Bergen University College, Bergen, Norway
\item \Idef{org36}Faculty of Mathematics, Physics and Informatics, Comenius University, Bratislava, Slovakia
\item \Idef{org37}Faculty of Nuclear Sciences and Physical Engineering, Czech Technical University in Prague, Prague, Czech Republic
\item \Idef{org38}Faculty of Science, P.J.~\v{S}af\'{a}rik University, Ko\v{s}ice, Slovakia
\item \Idef{org39}Frankfurt Institute for Advanced Studies, Johann Wolfgang Goethe-Universit\"{a}t Frankfurt, Frankfurt, Germany
\item \Idef{org40}Gangneung-Wonju National University, Gangneung, South Korea
\item \Idef{org41}Gauhati University, Department of Physics, Guwahati, India
\item \Idef{org42}Helsinki Institute of Physics (HIP), Helsinki, Finland
\item \Idef{org43}Hiroshima University, Hiroshima, Japan
\item \Idef{org44}Indian Institute of Technology Bombay (IIT), Mumbai, India
\item \Idef{org45}Indian Institute of Technology Indore, Indore (IITI), India
\item \Idef{org46}Inha University, Incheon, South Korea
\item \Idef{org47}Institut de Physique Nucl\'eaire d'Orsay (IPNO), Universit\'e Paris-Sud, CNRS-IN2P3, Orsay, France
\item \Idef{org48}Institut f\"{u}r Informatik, Johann Wolfgang Goethe-Universit\"{a}t Frankfurt, Frankfurt, Germany
\item \Idef{org49}Institut f\"{u}r Kernphysik, Johann Wolfgang Goethe-Universit\"{a}t Frankfurt, Frankfurt, Germany
\item \Idef{org50}Institut f\"{u}r Kernphysik, Westf\"{a}lische Wilhelms-Universit\"{a}t M\"{u}nster, M\"{u}nster, Germany
\item \Idef{org51}Institut Pluridisciplinaire Hubert Curien (IPHC), Universit\'{e} de Strasbourg, CNRS-IN2P3, Strasbourg, France
\item \Idef{org52}Institute for Nuclear Research, Academy of Sciences, Moscow, Russia
\item \Idef{org53}Institute for Subatomic Physics of Utrecht University, Utrecht, Netherlands
\item \Idef{org54}Institute for Theoretical and Experimental Physics, Moscow, Russia
\item \Idef{org55}Institute of Experimental Physics, Slovak Academy of Sciences, Ko\v{s}ice, Slovakia
\item \Idef{org56}Institute of Physics, Academy of Sciences of the Czech Republic, Prague, Czech Republic
\item \Idef{org57}Institute of Physics, Bhubaneswar, India
\item \Idef{org58}Institute of Space Science (ISS), Bucharest, Romania
\item \Idef{org59}Instituto de Ciencias Nucleares, Universidad Nacional Aut\'{o}noma de M\'{e}xico, Mexico City, Mexico
\item \Idef{org60}Instituto de F\'{\i}sica, Universidad Nacional Aut\'{o}noma de M\'{e}xico, Mexico City, Mexico
\item \Idef{org61}iThemba LABS, National Research Foundation, Somerset West, South Africa
\item \Idef{org62}Joint Institute for Nuclear Research (JINR), Dubna, Russia
\item \Idef{org63}Konkuk University, Seoul, South Korea
\item \Idef{org64}Korea Institute of Science and Technology Information, Daejeon, South Korea
\item \Idef{org65}KTO Karatay University, Konya, Turkey
\item \Idef{org66}Laboratoire de Physique Corpusculaire (LPC), Clermont Universit\'{e}, Universit\'{e} Blaise Pascal, CNRS--IN2P3, Clermont-Ferrand, France
\item \Idef{org67}Laboratoire de Physique Subatomique et de Cosmologie, Universit\'{e} Grenoble-Alpes, CNRS-IN2P3, Grenoble, France
\item \Idef{org68}Laboratori Nazionali di Frascati, INFN, Frascati, Italy
\item \Idef{org69}Laboratori Nazionali di Legnaro, INFN, Legnaro, Italy
\item \Idef{org70}Lawrence Berkeley National Laboratory, Berkeley, CA, United States
\item \Idef{org71}Lawrence Livermore National Laboratory, Livermore, CA, United States
\item \Idef{org72}Moscow Engineering Physics Institute, Moscow, Russia
\item \Idef{org73}National Centre for Nuclear Studies, Warsaw, Poland
\item \Idef{org74}National Institute for Physics and Nuclear Engineering, Bucharest, Romania
\item \Idef{org75}National Institute of Science Education and Research, Bhubaneswar, India
\item \Idef{org76}Niels Bohr Institute, University of Copenhagen, Copenhagen, Denmark
\item \Idef{org77}Nikhef, National Institute for Subatomic Physics, Amsterdam, Netherlands
\item \Idef{org78}Nuclear Physics Group, STFC Daresbury Laboratory, Daresbury, United Kingdom
\item \Idef{org79}Nuclear Physics Institute, Academy of Sciences of the Czech Republic, \v{R}e\v{z} u Prahy, Czech Republic
\item \Idef{org80}Oak Ridge National Laboratory, Oak Ridge, TN, United States
\item \Idef{org81}Petersburg Nuclear Physics Institute, Gatchina, Russia
\item \Idef{org82}Physics Department, Creighton University, Omaha, NE, United States
\item \Idef{org83}Physics Department, Panjab University, Chandigarh, India
\item \Idef{org84}Physics Department, University of Athens, Athens, Greece
\item \Idef{org85}Physics Department, University of Cape Town, Cape Town, South Africa
\item \Idef{org86}Physics Department, University of Jammu, Jammu, India
\item \Idef{org87}Physics Department, University of Rajasthan, Jaipur, India
\item \Idef{org88}Physik Department, Technische Universit\"{a}t M\"{u}nchen, Munich, Germany
\item \Idef{org89}Physikalisches Institut, Ruprecht-Karls-Universit\"{a}t Heidelberg, Heidelberg, Germany
\item \Idef{org90}Politecnico di Torino, Turin, Italy
\item \Idef{org91}Purdue University, West Lafayette, IN, United States
\item \Idef{org92}Pusan National University, Pusan, South Korea
\item \Idef{org93}Research Division and ExtreMe Matter Institute EMMI, GSI Helmholtzzentrum f\"ur Schwerionenforschung, Darmstadt, Germany
\item \Idef{org94}Rudjer Bo\v{s}kovi\'{c} Institute, Zagreb, Croatia
\item \Idef{org95}Russian Federal Nuclear Center (VNIIEF), Sarov, Russia
\item \Idef{org96}Russian Research Centre Kurchatov Institute, Moscow, Russia
\item \Idef{org97}Saha Institute of Nuclear Physics, Kolkata, India
\item \Idef{org98}School of Physics and Astronomy, University of Birmingham, Birmingham, United Kingdom
\item \Idef{org99}Secci\'{o}n F\'{\i}sica, Departamento de Ciencias, Pontificia Universidad Cat\'{o}lica del Per\'{u}, Lima, Peru
\item \Idef{org100}Sezione INFN, Bari, Italy
\item \Idef{org101}Sezione INFN, Bologna, Italy
\item \Idef{org102}Sezione INFN, Cagliari, Italy
\item \Idef{org103}Sezione INFN, Catania, Italy
\item \Idef{org104}Sezione INFN, Padova, Italy
\item \Idef{org105}Sezione INFN, Rome, Italy
\item \Idef{org106}Sezione INFN, Trieste, Italy
\item \Idef{org107}Sezione INFN, Turin, Italy
\item \Idef{org108}SSC IHEP of NRC Kurchatov institute, Protvino, Russia
\item \Idef{org109}SUBATECH, Ecole des Mines de Nantes, Universit\'{e} de Nantes, CNRS-IN2P3, Nantes, France
\item \Idef{org110}Suranaree University of Technology, Nakhon Ratchasima, Thailand
\item \Idef{org111}Technical University of Split FESB, Split, Croatia
\item \Idef{org112}The Henryk Niewodniczanski Institute of Nuclear Physics, Polish Academy of Sciences, Cracow, Poland
\item \Idef{org113}The University of Texas at Austin, Physics Department, Austin, TX, USA
\item \Idef{org114}Universidad Aut\'{o}noma de Sinaloa, Culiac\'{a}n, Mexico
\item \Idef{org115}Universidade de S\~{a}o Paulo (USP), S\~{a}o Paulo, Brazil
\item \Idef{org116}Universidade Estadual de Campinas (UNICAMP), Campinas, Brazil
\item \Idef{org117}University of Houston, Houston, TX, United States
\item \Idef{org118}University of Jyv\"{a}skyl\"{a}, Jyv\"{a}skyl\"{a}, Finland
\item \Idef{org119}University of Liverpool, Liverpool, United Kingdom
\item \Idef{org120}University of Tennessee, Knoxville, TN, United States
\item \Idef{org121}University of Tokyo, Tokyo, Japan
\item \Idef{org122}University of Tsukuba, Tsukuba, Japan
\item \Idef{org123}University of Zagreb, Zagreb, Croatia
\item \Idef{org124}Universit\'{e} de Lyon, Universit\'{e} Lyon 1, CNRS/IN2P3, IPN-Lyon, Villeurbanne, France
\item \Idef{org125}V.~Fock Institute for Physics, St. Petersburg State University, St. Petersburg, Russia
\item \Idef{org126}Variable Energy Cyclotron Centre, Kolkata, India
\item \Idef{org127}Vestfold University College, Tonsberg, Norway
\item \Idef{org128}Warsaw University of Technology, Warsaw, Poland
\item \Idef{org129}Wayne State University, Detroit, MI, United States
\item \Idef{org130}Wigner Research Centre for Physics, Hungarian Academy of Sciences, Budapest, Hungary
\item \Idef{org131}Yale University, New Haven, CT, United States
\item \Idef{org132}Yonsei University, Seoul, South Korea
\item \Idef{org133}Zentrum f\"{u}r Technologietransfer und Telekommunikation (ZTT), Fachhochschule Worms, Worms, Germany
\end{Authlist}
\endgroup

\end{document}